\documentclass[english,aps,prb,floatfix,twocolumn,showpacs]{revtex4}
\usepackage{graphicx}

\begin{document}

\title{Electronic structure of V$_{2}$O$_{3}$ : Wannier orbitals from LDA-$N$MTO calculations}
\author{T. Saha-Dasgupta,$^{1,2}$ O. K. Andersen,$^{1}$ J. Nuss,$^{1}$ A. I.
Poteryaev,$^{3,4}$ A. Georges,$^{3}$ and A. I. Lichtenstein.$^{5}$}
\affiliation{$^{1}$Max-Planck Institut f\"{u}r Festk\"{o}rperforschung, Heisenbergstrasse
1, D-70569, Stuttgart, Germany\\
$^{2}$S.N.Bose Center for Basic Sciences, Salt Lake, Kolkata 700098, India\\
$^{3}$Centre de Physique Th\'{e}orique, Ecole Polytechnique, F91128
Palaiseau CEDEX, France\\
$^{4}$ Institute of Metal Physics, S. Kovalevskaya 18, GSP-170, 620041, Ekaterinburg, Russia \\
$^{5}$I. Institut f\"{u}r Theoretische Physik, Universit\"{a}t Hamburg,
Jungiusstrasse 9, D-20355 Hamburg, Germany}
\pacs{71.20.-b, 71.15.Ap, 71.20.Be}
\date{\today }

\begin{abstract}
Using muffin-tin orbital (MTO) based {\it N}MTO-downfolding
procedure within the framework of local density approximation, we
construct the Wannier orbitals for the $t_{2g}$  manifold of bands
in V$_{2}$O$_{3}$ in the paramagnetic phase. The real space
representation of the one-electron Hamiltonian in the constructed
Wannier function basis shows that, contrary to the popular belief,
the in-plane hopping interactions are as important as the vertical
pair hopping. Following the language of Di Matteo {\it et.al.}
[Phys. Rev. B {\bf 65} 054413 (2002)], this implies, the problem 
of V$_{2}$O$_{3}$ falls in the atomic regime rather than in the
molecular regime. We have also repeated our construction
procedure in the low temperature monoclinic phase, for which the
changes in hopping interactions are found not to be dramatic. 

\end{abstract}

\maketitle

\section{Introduction}

V$_{2}$O$_{3}$ has been in focus of attention since 1969 when its unusual
phase diagram was discovered \cite{mcwhan}. At low temperature, pure V$_{2}$O%
$_{3}$ is an antiferromagnetic insulator (AFI) with a monoclinic, slightly
distorted corundum structure, a complicated magnetic order, a moment of
1.2\thinspace $\mu _{B},$ and a gap of 0.66\thinspace eV \cite{afm}. At $%
T_{N}$=154\thinspace K it transforms to a corundum-structured, paramagnetic
metal. Upon substituting V by Ti or by application of pressure, the Ne\'{e}l
temperature decreases and the antiferromagnetic phase vanishes above 5\% Ti.
Substitution of V by Cr, on the other hand, causes the Ne\'{e}l temperature
to increase and reach 180 K for 1.8\% Cr. For higher concentrations, $T_{N}$
stays constant \emph{and} the transition is to a paramagnetic \emph{%
insulator.} For Cr concentrations between 0.5 and 1.8\% there is a \emph{%
second} phase transition, which upon increasing temperature, or Cr
concentration, is from a paramagnetic metal (PM) to a paramagnetic insulator (PI).
This transition is isostructural, ends at a critical point, $\left(
T_{c},y_{c}\right) =\left( 400\,\mathrm{K},\,0.5\%\text{ }\mathrm{Cr}\right)
,$ and has been considered the classic example of a Mott-Hubbard transition.

In the high-temperature corundum structure (FIG.\ \ref{structure}), all vanadium ions are
equivalent and surrounded by nearly perfect oxygen octahedra. Since the
covalent O-V $pd\sigma $ interaction is stronger than the $pd\pi $
interaction, the more antibonding V\thinspace $d$-like $e_{g}$ level\cite{note} lies
above the less antibonding V\thinspace $d$-like $t_{2g}$ level and, as a
consequence, the electronic configuration of V$_{2}$O$_{3}$ is V\thinspace $%
t_{2g}^{2}.$ Now, the three-fold degenerate $t_{2g}$ level is split into an
upper $a_{1g}$ and a lower, doubly degenerate $e_{g}^{\pi }$ level by a
trigonal distortion, which mainly consists of a slight displacement of the
vanadium ions along the vertical three-fold axis, away from the centers of
their octahedra, so that the distance between a \emph{vertical vanadium pair}
(V 4-1 or 2-\b{5} in FIG.\ \ref{structure}) is slightly longer than the distance between
the centers of the two octahedra.

There have been many attempts to explain these metal-insulator transitions
and the spin structure of the antiferromagnetic insulating phase. The
careful analysis presented in 1978 by Castellani, Natoli, and Ranninger
(CNR) \cite{cast} resulted in a model which remained undisputed
for over twenty years: Since the $a_{1g}$ orbitals have $d_{3z^{2}-1}$
character and point towards each other, yielding a strong $dd\sigma $-like
hopping integral, these orbitals on each vertical pair form bonding and
antibonding levels which are split by more than twice the $a_{1g}$-$%
e_{g}^{\pi }$ crystal-field splitting and by more than the on-site Coulomb
interaction. As a consequence, \emph{one} electron per vanadium is used to
form a spin-singlet, chemical bond between a vertical pair. The \emph{other}
electron enters the doubly degenerate, localized $e_{g}^{\pi }$ orbitals.
Since the integrals for hopping from- and between $e_{g}^{\pi }$ orbitals
are relatively small, the on-site Coulomb repulsion leads to an $S$=1/2
state and may \emph{order} the occupied $e_{g}^{\pi }$ orbitals in a way
consistent with the observed spin structure of the low-temperature
antiferromagnetic insulator. In this structure, the spins on vanadium pairs
in the $z$ and $x$ directions are aligned ferromagnetically, and those on
pairs in the other two directions are aligned antiferromagnetically. This
requires an orbital order in which an integral for hopping in the $x$ and $z$
directions between occupied $e_{g}^{\pi }$ orbitals is considerably smaller
than between occupied and unoccupied $e_{g}^{\pi }$ orbitals. This was
reviewed and discussed by Rice\cite{rice} using Kugel-Khomskii's general description 
\cite{KK} of the coupling between orbital and spin degrees of
freedom. The CNR model also led to a half-filled, one-band (the lowest $%
e_{g}^{\pi }$-band) Hubbard Hamiltonian to serve as the simplest possible
electronic model for V$_{2}$O$_{3}$. This model was solved by Rozenberg 
\textit{et. al.}\cite{rozen} using the dynamical mean-field approximation
(DMFT)\cite{dmft-rev} and found to describe the metal-insulator transition.
However, the polarized x-ray absorption experiment of Park {\it et. al.} 
\cite{park}, corroborated with multiplet calculations showed that 
V 3$d^{2}$ ions are in the high spin ($S=1$) state rather than in 
$S=1/2$ state and the orbital occupation, which is different in different
phases, is an admixture of $e_{g}^{\pi}e_{g}^{\pi}$ states
with $e_{g}^{\pi}a_{1g}$ configurations. From that, they concluded
that neither V$_{2}$O$_{3}$ problem can be mapped onto a
single-band Hubbard model, nor the projecting out of $a_{1g}$
orbitals by means of molecular orbital formation as was done
by  CNR\cite{cast} is justified. Ezhov {\it et. al.}'s 
calculations \cite{ezhov} within the local density approximation (LDA)+U 
scheme showed the importance of Hund's rule exchange giving rise to 
$S=1$ model but with only $e_{g}^{\pi}$ occupancy implying no
orbital ordering. However, they succeeded in correctly predicting
the low-temperature magnetic structure which was attributed to be
stabilized by the monoclinic distortion. Nevertheless, the issue
associated with the orbital ordering remained which apparently
showed up its presence in several different experimental observations
\cite{expt}. To reconcile the $S=1$ and the orbital ordering aspect,
Mila {\it et. al.} \cite{mila} and Di Matteo {\it et. al.}
\cite{matteo} subsequently proposed two different correlated model
of $c-axis$ pair states incorporating dynamical mixing of  $e_{g}^{\pi}e_{g}^{\pi}$
and $e_{g}^{\pi}a_{1g}$  states with $S=1$ spin configuration
on each of the sites. In recent years, combined with the LDA, DMFT 
calculations have been carried out\cite{dmft}.

While, it is now generally accepted, a realistic theory of
V$_{2}$O$_{3}$ must take into account the complicated electronic
structure of the system, there has been no serious attempt to realistic
modeling of the electronic structure of V$_{2}$O$_{3}$ since the
early work of CNR\cite{cast} which was crude in 
its various approximations and was partly semi-empirical. The
starting point of several of the many-body model-based calculations
seem to be the vertical V-V pair model which is
considered to be the predominant building blocks $-$ the validity
of such assumptions need to be re-examined in the context of
accurate tight-binding (TB) modeling of V$_{2}$O$_{3}$. 

In recent years MTO based $N-th$ order MTO method,
namely {\it N}MTO method\cite{nmto1,nmto2} has been introduced and implemented. 
The method goes beyond the scope of the standard linear MTO (LMTO) method, in defining 
an energetically accurate basis set with a consistent description throughout 
the space of MT spheres and the interstitial. An important feature of the 
{\it N}MTO method is the so-called {\it downfolding} technique which provides an 
useful way to derive few-orbital Hamiltonians starting from complicated full 
LDA Hamiltonian by integrating out degrees of freedom not-of-interest. This 
procedure naturally takes into account the renormalization effect due of the 
integrated-out orbitals by defining energy-selective, effective
orbitals which serve as the Wannier or Wannier-like orbitals for 
the few-orbital Hamiltonian in {\it downfolded}
representation. The method provides a first-principles way of deriving the 
single-particle model Hamiltonian and direct generation of Wannier 
functions without any fitting procedure giving rise to 
an unique scheme that has the deterministic nature of first-principles
calculations added to the simplicity of model Hamiltonian
approaches.  The method has proved to be extremely
successful in deriving model Hamiltonians for systems 
such as  high-Tc cuprates \cite{app1}, double perovskites \cite{app2}, 
low-dimensional quantum spin systems \cite{app3}.
This approach of direct generation of Wannier functions may be 
contrasted to that of construction of Wannier functions
out of the calculated Bloch functions. Recently Anisimov \textit{et al.} implemented a similar method
for use in LDA+DMFT \cite{AnisimovFull} and Solovyev proposed a general
LMTO-based procedure for constructing effective lattice fermion models\cite%
{Solovyev06}. With other local-orbital basis sets, somewhat similar
techniques can be used \cite{KuWei}, but in case not all basis functions are
well localized, e.g. for the set of bare LMTOs\cite{LMTO75}, the Wannier
functions obtained for the correlated bands may not be sufficently localized
for the corresponding on-site-$U$-Hamiltonian to be realistic. For those
cases, more complicated procedures for obtaining for instance those Wannier
functions which minimize the spread,\cite{mv} $\left\langle \left\vert 
\mathbf{r}-\left\langle \mathbf{r}\right\rangle \right\vert
^{2}\right\rangle ,$ or those which maximize the Coulomb self-energy, has
been used.\cite{Albers02,Lechermann06}

In this paper, we aim
to provide an accurate tight-binding description of 
V$_{2}$O$_{3}$ by constructing the Wannier-like functions for the $t_{2g}$ Hamiltonian
employing the {\it N}MTO methodology. In this context, this methodology has recently been applied \cite{ilya} 
for tight-binding modeling of $a_{1g}$ bands of ferromagnetic LDA+U 
calculations to investigate the role of vertical pair from band-structure 
point of view.
In those specific calculations, primary interest was to estimate
the $c-axis$ intra-pair hopping matrix element compared to inter-pair hopping
matrix elements for $a_{1g}$ bands and therefore, LDA+U rather than
LDA was chosen as the basis of calculations, which provides nice separation of
$a_{1g}$ and $e_{g}^{\pi}$ bands. However, from point of view of 
input to realistic many-body calculations it is more suitable and
preferable to start with LDA-derived 
Hamiltonians and the full $t_{2g}$ Hamiltonian since the $a_{1g}$ and 
$e_{g}^{\pi}$ states are both equally important as seen in experiment and
the hybridization effect between $a_{1g}$ and 
$e_{g}^{\pi}$ should be taken in account.
In the present work, we have therefore chosen LDA as the basis
of our calculations. 
The present paper in that respect, should be considered
as a more detail paper for the tight-binding modeling of V$_{2}$O$_3$
system. The tight-binding parameters derived in this paper in a 
rigorous, first-principles manner will be useful as an input to
many-body variational calculation like that of Di Matteo {\it et. al.} 
\cite{matteo} as will be discussed in section III D. The parameters
can also be used for the many-body LDA+DMFT calculations. The LDA+DMFT calculations
report in Ref. \cite{v2o3II}, has been carried out using the
{\it N}MTO Wannier function implementation of LDA, presented in this paper.
These calculations\cite{v2o3II} showed the importance of correlation
assisted dehybridization of $a_{1g}$ and $e_{g}^{\pi}$ in the description
of the correlated electronic structure of V$_{2}$O$_3$ and its metal-insulator
transition. Calculations within such LDA+DMFT framework
has been also used to explore the comparison of doping, temperature and
pressure route to metal-insulator transition in V$_{2}$O$_3$\cite{xas}
and to study multi-orbital effects in optical properties of V$_{2}$O$_{3}$\cite{silke,giorgio}.

In the following, in section II we discuss the crystal structure and orbital
symmetry aspects in V$_{2}$O$_{3}$. 
Section III involves description of the results and discussion. This
section is divided into several sub-sections. In the sub-section A we
present the high-energy part of the LDA band-structure of rhombohedral,
undoped V$_{2}$O$_{3}$ in ambient pressure. In the sub-section B we explain 
how downfolding within the {\it N}MTO method can be used to construct truly 
minimal basis sets which pick out selectively O-$p$, V-$t_{2g}$, V$e_{g}$ or 
V-$s$ bands. We also present the {\it N}MTOs - the members of such truly minimal 
sets of V-$t_{2g}$ and V$e_{g}$. With this tool at hand, in sub-section C we zoom in
on the LDA $t_{2g}$ bands, where due to lowering of symmetry induced by
trigonal distortion we switch on from the $t_{2g}$ representation to more
appropriate $a_{1g} - e_{g}^{\pi}$ representation. We present the minimal
set constructed out of $a_{1g}$ and $e_{g}^{\pi}$'s, their Wannier functions,
the hopping integrals and the comparison with existing results. In 
sub-section D, we discuss the validity of vertical pair model and
the molecular orbital based approaches
in light of {\it N}MTO derived hopping integrals. 
We restricted our study to high temperature
paramagnetic phase of pure V$_{2}$O$_{3}$ in rhombohedral, corundum structure until
the sub-section E, where for the sake of
completeness of our study, we also discuss the hopping 
integrals in the Cr-doped V$_{2}$O$_{3}$ (sub-section E) and in the 
low-temperature monoclinic structure (sub-section F). 
Finally, we conclude in section IV with summary and outlook. 
The essential
details about the {\it N}MTO method, which will be used for construction of
localized Wannier orbitals and truly minimal, downfolded basis sets can
be found in the Appendix A.


\section{Crystal structure and symmetry:}

As mentioned in the introduction, the high temperature paramagnetic 
phase of V$_{2}$O$_{3}$ has the corundum structure. This structure 
consists of hexagonal packing of the oxygen atoms, and 
the vanadium atoms occupying 2/3 of the octahedral cation sites.
The basic features of the corundum structure is shown in FIG.\ \ref{structure}. 
The immediate surrounding of the V atoms provided by oxygen atoms 
has approximate octahedral symmetry. The VO$_{6}$ octahedra
face share along the vertical direction forming V-V vertical bonds, 
while they edge share forming layers of honeycomb lattice, giving
rise to three-dimensional network with overall rhombohedral symmetry.
In the resulting corundum structure, which has R$\bar{3}$c space group
symmetry, each primitive rhombohedral unit cell contains two 
V$_{2}$O$_{3}$ formula units while the non-primitive hexagonal unit 
cell contains six V$_{2}$O$_{3}$ formula units. The experimentally
determined structure \cite{struc} of ambient pressure, pure V$_{2}$O$_{3}$,
in the hexagonal setting, with lattice constants $a_H$=4.952 $\AA$ and $c_H$=14.003 $\AA$, 
and V and O atoms occupying the Wyckoff 
positions (12c) and (18e) with internal parameters $z_V$ = 0.34630
and $x_O$ = 0.31164, yield V-O bond lengths in the range 1.971 
$\AA$- 2.049 $\AA$. The nearest-neighbor V-V distances within 
the hexagonal layers are 2.882 $\AA$ while that along the vertical 
direction is about 6 $\%$  shorter (2.697 $\AA$). As is evident from 
the structural figure as well as from the internal parameter value, 
the V atoms are displaced from their {\it ideal} positions where V atoms
in the hexagonal layers would have been co-planer. This
displacement causes V atoms to move away from the center of the
octehedra, giving rise to three long and three short V-O bonds. The arrangement
of V atoms along the hexagonal z-axis can be derived from an ideal chain
structure by introducing vacancies at every third site.

\begin{widetext}
\begin{center}
\begin{figure}
\includegraphics[width=16cm,keepaspectratio]{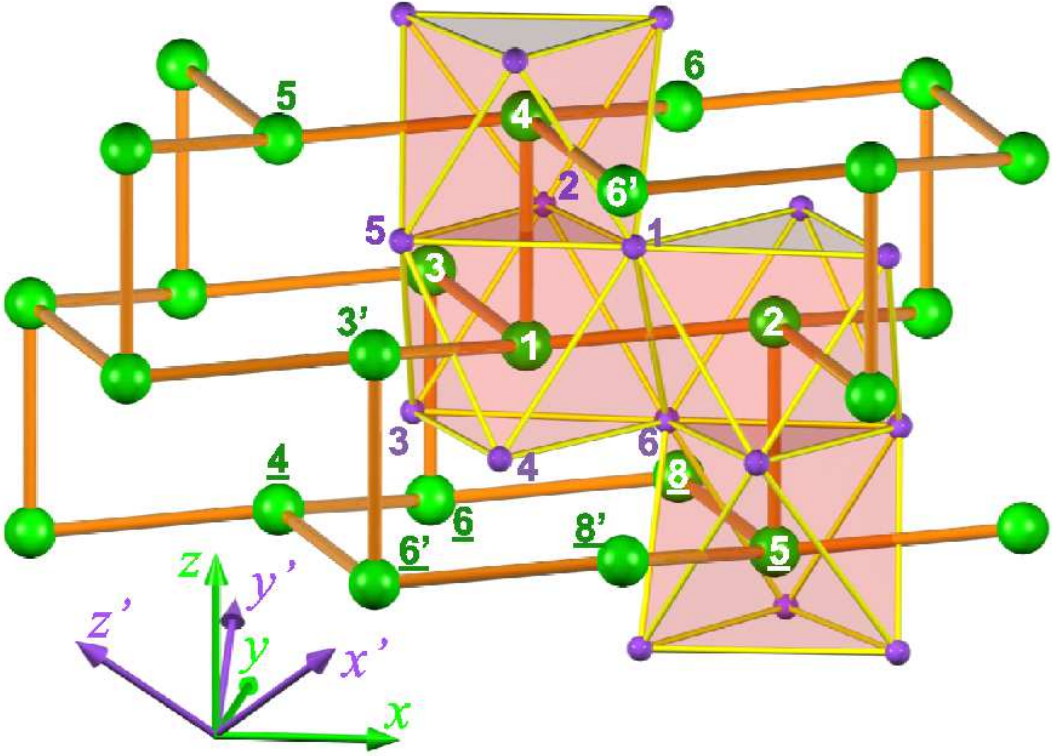}
\caption{(Color online) The crystal structure of V$_{2}$O$_{3}$ in the
  high-temperature paramagnetic phase. The larger (green) circles
  indicate the V atoms. The smaller (violet) circles surrounding
  the V atoms are oxygens, showing the octahedral
  co-ordination. The VO$_{6}$ octahedra face-share along the
  vertical direction while they edge-share within the hexagonal
  layers. The unprimed and primed co-ordinate systems represent the
  rhombohedral co-ordinate system ($z$-axis pointing along the
  vertical V-V bond and $x$-axis chosen as the projection of V1-V2 in
  the $xy$ plane) and the oxygen-based octahedral co-ordinate system
  ($z^{'}$-axis pointing along O6-O5, $x{'}$-axis pointing along
  O3-O1) respectively.}
\label{structure}
\end{figure}
\end{center}
\end{widetext}

\begin{figure}
\includegraphics[width=7cm,keepaspectratio]{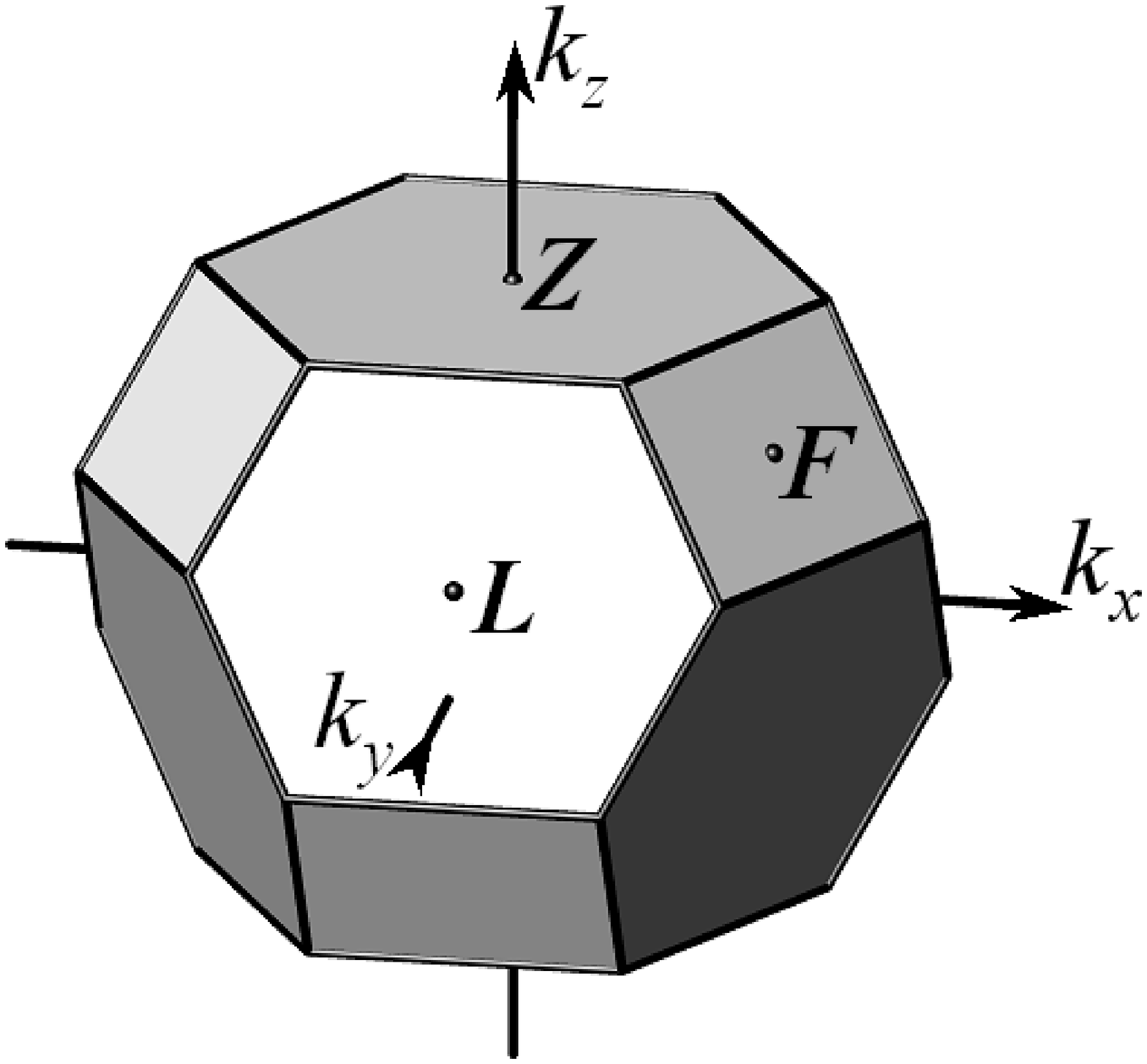}
\caption{The rhombohedral BZ showing the high-symmetry points.}
\label{bz}
\end{figure}

The results reported in the following, for the high-temperature
structure, are all carried out with rhombohedral
unit cell defined by the primitive lattice translations in a Cartesian
system with $z$-axis pointing along the vertical V-V bond and the $x$-axis
chosen as the projection of V1-V2 onto the $xy$ plane:

\[
\left[ 
\begin{array}{c}
{\bf T}_{1} \\ 
{\bf T}_{2} \\ 
{\bf T}_{3}%
\end{array}%
\right] ={\rm a}\left[ 
\begin{array}{ccc}
1 & 0 & \frac{{\rm c}}{{\rm a}} \\ 
-\frac{1}{2} & \frac{\sqrt{3}}{2} & \frac{{\rm c}}{{\rm a}} \\ 
-\frac{1}{2} & -\frac{\sqrt{3}}{2} & \frac{{\rm c}}{{\rm a}}%
\end{array}%
\right] .
\]%

where c/a = 1.633 and a = 2.859 $\AA$. The 3-fold axis pointing along
the $z-$direction are given by ${\bf T}_{1} + {\bf T}_{2} + {\bf T}_{3}$. The
primitive translations in the reciprocal lattice, $
\left[ 
\begin{array}{ccc}
{\bf G}_{1} & {\bf G}_{2} & {\bf G}_{3}%
\end{array}%
\right] $
defined as $2\pi \left[ \begin{array}{c}
{\bf T}_{1} \\ 
{\bf T}_{2} \\ 
{\bf T}_{3}%
\end{array}%
\right] ^{-1}$
are given by
\[
\frac{2\pi }{{\rm a}}\frac{2}{3}\left[ 
\begin{array}{ccc}
1 & -\frac{1}{2} & -\frac{1}{2} \\ 
0 & \frac{\sqrt{3}}{2} & -\frac{\sqrt{3}}{2} \\ 
\frac{\alpha }{2} & \frac{\alpha }{2} & \frac{\alpha }{2}%
\end{array}%
\right]. 
\]%
where $\alpha$ = (c/a)$^{-1}$. \\ \\

The high-symmetry points on the Brillouin zone (BZ) as shown in
FIG.\ \ref{bz} 
are given by, \\ \\
2 points on the 3-fold axis and at the center of regular hexagonal 
faces of the BZ : \\
\[
{\bf Z=\pm }\frac{1}{2}\left( {\bf G}_{1}{\bf +G}_{2}{\bf +G}_{3}\right)
=\pm \frac{2\pi }{{\rm a}}\left[ 
\begin{array}{c}
0 \\ 
0 \\ 
\frac{\alpha }{2}%
\end{array}%
\right] ,
\]%
6 points at the center of hexagonal faces of BZ with 2 short and 4 long edges:
\begin{eqnarray*}
{\bf L} &=\pm& \frac{1}{2}{\bf G}_{1}=\pm \frac{2\pi }{{\rm a}}\frac{1}{3}\left[
\begin{array}{c}
1 \\ 
0 \\ 
\frac{\alpha }{2}%
\end{array}%
\right] ,\;\;{\bf \pm }\frac{1}{2}{\bf G}_{2}=\pm \frac{2\pi }{{\rm a}}\frac{%
1}{3}\left[ 
\begin{array}{c}
-\frac{1}{2} \\ 
\frac{\sqrt{3}}{2} \\ 
\frac{\alpha }{2}%
\end{array}%
\right] ,\\
& & \;\;{\bf \pm }\frac{1}{2}{\bf G}_{3}=\pm \frac{2\pi }{{\rm a}}\frac{%
1}{3}\left[ 
\begin{array}{c}
-\frac{1}{2} \\ 
-\frac{\sqrt{3}}{2} \\ 
\frac{\alpha }{2}%
\end{array}%
\right] 
\end{eqnarray*}
and 6 points at the center of rectangular faces:
\[
{\bf F=}\pm \frac{1}{2}\left( {\bf G}_{2}{\bf +G}_{3}\right) =\pm \frac{2\pi 
}{{\rm a}}\frac{1}{3}\left[ 
\begin{array}{c}
-1 \\ 
0 \\ 
\alpha 
\end{array}%
\right], {\bf \;\;\pm }\frac{1}{2}\left( {\bf G}_{1}{\bf +G}_{2}\right) \]
\[ =%
{\bf \pm }\frac{2\pi }{{\rm a}}\frac{1}{3}\left[ 
\begin{array}{c}
\frac{1}{2} \\ 
\frac{\sqrt{3}}{2} \\ 
\alpha 
\end{array}%
\right] ,\;\;{\bf \pm }\frac{1}{2}\left( {\bf G}_{1}{\bf +G}_{3}\right) =\pm 
\frac{2\pi }{{\rm a}}\frac{1}{3}\left[ 
\begin{array}{c}
\frac{1}{2} \\ 
-\frac{\sqrt{3}}{2} \\ 
\alpha 
\end{array}%
\right] 
\]%

The perfect octahedral crystal field surrounding of the V ions split the 3d 
energy levels into two-fold degenerate $e_{g}$ levels, 
d$_{3{z^{'}}^{2}-1}$, d$_{{x^{'}}^{2}-{y^{'}}^{2}}$ and three-fold 
degenerate $t_{2g}$ levels, d$_{x^{'}y^{'}}$, d$_{x^{'}z^{'}}$, 
d$_{y^{'}z^{'}}$, where the primed co-ordinate system refers to the octahedral
co-ordinate system with $x^{'},y^{'}$ and $z^{'}$ pointing along 
O3-O1, O4-O2 and O6-O5 [see FIG.\ \ref{structure}]. 
However, the oxygen octahedra surrounding of the V ion
in V$_{2}$O$_{3}$ is not quite perfect, but has the trigonal distortion. This
trigonal distortion of the octahedral environment of the V site and 
influence of non-cubic arrangement of more distant V ions
in the lattice lowers the
symmetry from octahedral $O_{h}$ group to $D_{3d}$ group, resulting into
further splitting of the $t_{2g}$ complex into singly degenerate $a_{1g}$
and two-fold degenerate $e_{g}^{\pi}$. Starting from three
congruent $t_{2g}$ orbitals,d$_{x^{'}y^{'}}$, d$_{x^{'}z^{'}}$, 
d$_{y^{'}z^{'}}$ which can be derived from each other by a
counter-clockwise rotation of $2 \pi/3$ around the three-fold
$z$-axis, the $a_{1g}$ and $e_{g}^{\pi}$ orbitals are generated as:
\vskip .1in

\begin{eqnarray*}
d_{m} & = & \frac{1}{\sqrt{3}}(d_{x^{'}z^{'}} + d_{x^{'}y^{'}}e^{2 \pi im/3}
+ d_{y^{'}z^{'}}e^{-2 \pi im/3}) 
\end{eqnarray*}
\begin{eqnarray*}
a_{1g} & : & m = 0, e_{g}^{\pi} : m = \pm 1 
\end{eqnarray*}
\begin{eqnarray*}
e_{g}^{\pi},1 & = &  \sqrt{2} Im \enskip d_{1}
=  1/\sqrt 2 (d_{x^{'}y^{'}} -  d_{y^{'}z^{'}}) 
\end{eqnarray*}
\begin{eqnarray}
e_{g}^{\pi},2 &=& \sqrt{2} Re\enskip d_{1}= \sqrt{2/3}d_{x^{'}z^{'}} - 1/\sqrt{6}(d_{x^{'}y^{'}}+ d_{y^{'}z^{'}})
\end{eqnarray}
\vskip .1in

\noindent

Further, transforming to
rhombohedral, unprimed co-ordinate system ${x,y,z}$, the three
orbitals transforming as the $a_{1g}$ and $e_{g}^{\pi}$
representations are given by \cite{cast},
\begin{eqnarray*}
a_{1g} & :& d_{3z^{2}-1} \\
e_{g}^{\pi},1 & = & \sqrt{2/3} d_{xy} + 1/\sqrt{3} d_{xz} \\
e_{g}^{\pi},2 & = & -\sqrt{2/3} d_{x^{2}-y^{2}} - 1/\sqrt{3} d_{yz} 
\end{eqnarray*}

The high-temperature Cr-doped V$_{2}$O$_{3}$, which is paramagnetic, insulator in
nature, retains the corundum crystal structure, with lattice constant expanding
to 4.998 $\AA$ and $c/a$ dropping to 2.78 for substitution of approximately 1 $\%$ 
Cr, compared to undoped, metallic phase discussed above. This causes expansion of
the vertical and all nearest-neighbor basal V-V bonds by 1.8-14 $\%$. Upon doping
with Cr, the Wyckoff positions of V and O also change. For (V$_{0.962}$Cr$_{0.038}$)$_{2}$O$_{3}$
they become $z_V$ = 0.34870
and $x_O$ = 0.30745, yielding V-O bond lengths in the range 1.976 
$\AA$- 2.061 $\AA$.

In the low-temperature AFI phase, the crystal structure is further distorted from the
corundum structure to monoclinic. This
distortion causes the tilting of the vertical V-V bond by 1.8$^{o}$ 
towards the positive side of the $x-$axis and breaks the three-fold rotational
symmetry, resulting into monoclinic crystal of symmetry I$_{2}/a$ 
with\cite{struc_mono} $a_m$ = 7.255 $\AA$, $b_m$ = 5.002 $\AA$, $c_m$ = 5.548
$\AA$, $\beta$ = 96.752$^{o}$  and four
formula weight per unit cell. As a consequence,
the vertical V-V bond length increases slightly from 2.697 $\AA$
to 2.745 $\AA$, one of the V-V bond (V1-V2) within the hexagonal
layer elongates to 2.986 from undistorted bond length of 2.882 $\AA$,
while the other two (V1-V3 and V1-V3$^{'}$) remain essentially
same with bond lengths 2.862 $\AA$ and 2.876 $\AA$. In the
low-temperature magnetic structure which is rather unusual, the vertical V1-V4
bond and the V1- V2 bond along the {\it x} axis becomes ferromagnetic while
the other two basal bonds become antiferromagnetic. Upon distortion,
the oxygen octahedra also becomes slightly skewed about the central V
atom, while the average V-O bond length remains
practically unaltered. In FIG.\ \ref{mono+hexa} we show the
low-temperature monoclinic structure together with high-temperature
corundum structure.
The low symmetry crystal field in the
monoclinic phase, further lifts the degeneracy between
two $e_{g}^{\pi}$ orbitals and mixes the  $a_{1g}$
and $e_{g}^{\pi}$ orbitals on the same site.

\begin{widetext}
\begin{center}
\begin{figure}
\includegraphics[width=12cm,keepaspectratio]{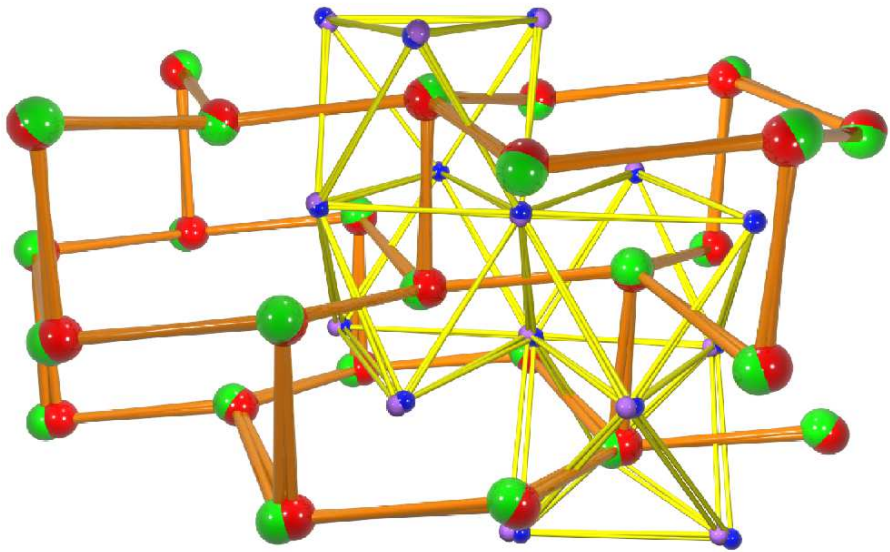}
\caption{(Color online) The crystal structure of V$_{2}$O$_{3}$ in the
  low-temperature monoclinic phase. For comparison, the
  high-temperature corundum structure is also shown in the same
  figure. The structures have been chosen to match at the central
  point of the vertical V1-V4 bond. As in FIG.\ \ref{structure} 
  the larger (red for corundum and green for monoclinic) circles
  indicate the V atoms. The smaller (blue for corundum and violet
  for monoclinic) circles surrounding the V atoms are oxygens.}
\label{mono+hexa}
\end{figure}
\end{center}
\end{widetext}


\section{Results and discussions}

\subsection{Corundum V$_{2}$O$_{3}$ : LDA band-structure}

FIG.\ \ref{fat} shows the LDA one-electron band-structure of corundum V$_{2}$O$_{3}$ in PM phase,
over an energy range of about 16 eV around the Fermi level (set as zero in the
figure). The bands are plotted  along the various symmetry
directions of the rhombohedral BZ, shown in FIG.\ \ref{bz}.
The results are
obtained with self-consistent potentials generated out of the
tight-binding LMTO calculation within the atomic sphere
approximation (ASA)\cite{lmto}. The details of the computation
may be found in Appendix B.
von Barth and Hedin parametrization \cite{bh} has been used for the LDA
exchange-correlation potential. The band-structure results
presented in the FIG.\ \ref{fat} are obtained with standard set of
nearly-orthonormal LMTO's, whose accuracy is good enough for
describing the high-energy features of the band-structure.
The bands are in good agreement with the
linear-augmented-plane-wave (LAPW) result of Mattheiss \cite{matt,note1}.

Plotted bands are the orbital-projected bands or the so-called {\it fat-band} in the
sense that the fatness of the bands in each panel is the
weight of the indicated orbital in the wave-function. The
co-ordinate system is chosen as that of the oxygen-based octahedral co-ordinate
system with the {\it z$^{'}$}-axis pointing along the O6-O5 direction and
{\it x$^{'}$}-axis pointing along the O3-O1 direction. As is seen,
the low-lying bands below -3 eV is predominantly of oxygen character. 
With the choice of octahedral co-ordinate system, the V $3d$ splits into $t_{2g}$ 
and $e_{g}$ manifolds. 12 $t_{2g}$-like bands (since there are 4 V atoms in the unit
cell of the primitive rhombohedral unit cell with 3 $t_{2g}$
orbitals on each V ion) lying lower in energy compared to $e_{g}$-like
bands cross the Fermi level, spanning an energy window from about
-1.5 eV to 1.5 eV. The crystal field split $e_{g}$-like bands
lye high up in energy from about 1.7 eV to 4 eV separated
from the $t_{2g}$ manifold by a small energy gap of about 0.2 eV. V-$s$
dominated states lye further high up in energy starting from about
4.5 eV. 

In the oxygen-projected band-structure, we notice, in addition to predominant fatness
associated with oxygen-dominated bands lying below -3eV, the fatness associated also
with V-$d$ dominated bands. Similarly in V-$d$ projected band-structure we notice the
presence of character in O-$p$ bands, which is born in by the V-$d$ $-$ O-$p$ hybridization.
It is this V-$d$ $-$ O-$p$ hybridization, that moves the V-$d$ and O-$p$ dominated states far apart
from each other with oxygen bands fully occupied and V-$d$ bands mostly empty. Due to the
different orientation, $e_{g}$ orbitals hybridize more strongly with O-$p$ forming directed
$pd{\sigma}$ bonds while the $t_{2g}$ orbitals bond less strongly giving rise to $pd{\pi}$
bonds. This is evident from the {\it fat-band} plots which shows that the upper part of the
V-$d$ bands - the energy region dominated by the V-$e_g$ and the lower part of the O-$p$ 
bands has the most mixing in. We also notice the significant hybridization between V-$s$
and O-$p$ degrees of freedom.

\begin{widetext}
\begin{center}
\begin{figure}
\includegraphics[width=16cm,keepaspectratio]{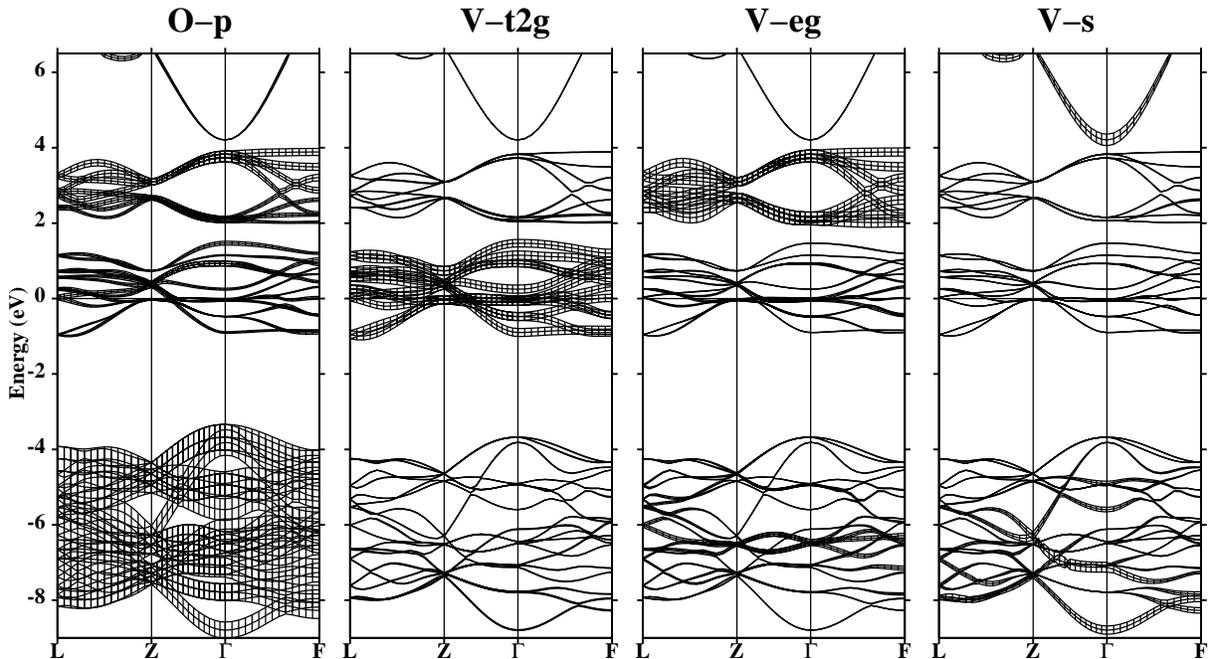}
\caption{LDA band-structure of V$_{2}$O$_{3}$ in the
  high-temperature corundum structure, plotted along the symmetry
  directions of the rhombohedral BZ. The BZ is shown in FIG.\
  \ref{bz}. The fatness associated with each band is proportional to
  the character of the orbital indicated at the top of each
  panel. Zero of the energy is set at the Fermi level.}
\label{fat}
\end{figure}
\end{center}
\end{widetext}

\subsection{Corundum V$_{2}$O$_{3}$: {\it downfolded} few-orbital band-structure and
Wannier-functions}

In the following, we demonstrate the application of 
{\it N}MTO-{\it downfolding} technique to produce truly minimal
basis sets which may be chosen to span selected bands with as few
basis orbitals as there are bands. This is illustrated by
constructing truly minimal basis sets for O-$p$, V-$t_{2g}$,
V-$e_{g}$ and V-$s$.
\begin{widetext}
\begin{center}
\begin{figure}
\includegraphics[width=17cm,keepaspectratio]{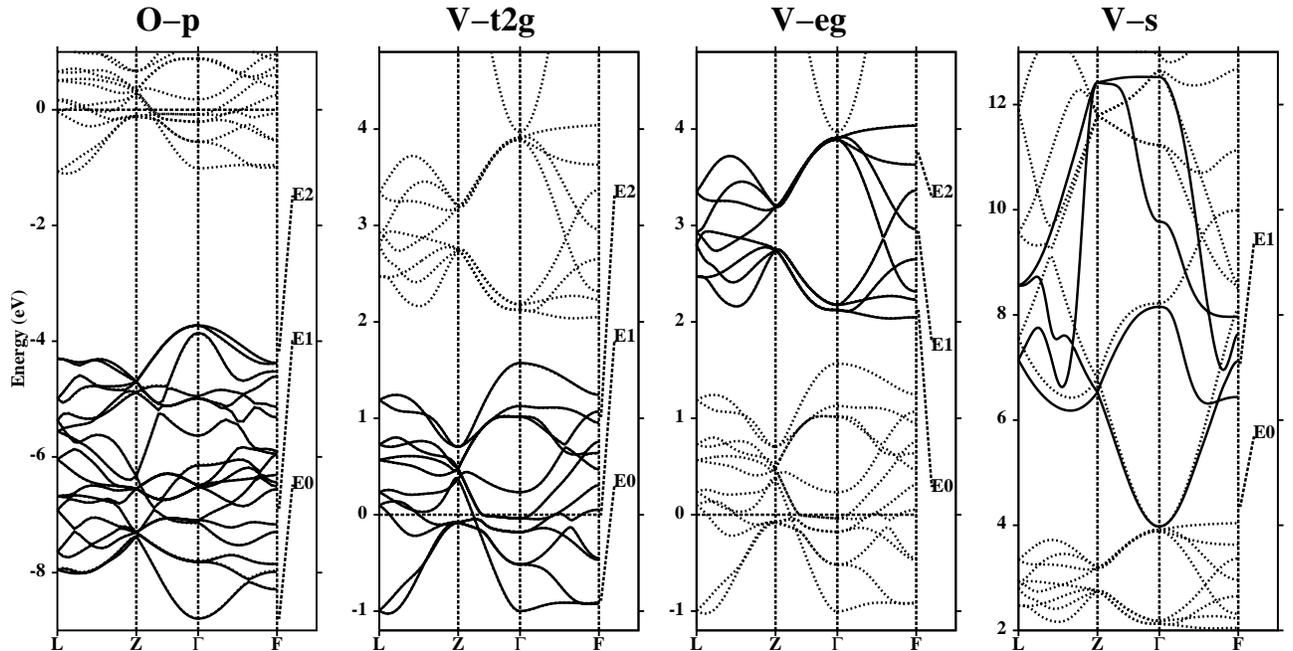}
\caption{LDA band structure of corundum-structured V$_{2}$O$_{3}$
  in various energy ranges. The solid lines in different panels
  show the bands obtained with the truly minimal (downfolded)
  O-$2p$, V-$t2_{2g}$, V-$e_g$ and V-$4s$ {\it N}MTO basis in comparison
  to those obtained with full {\it N}MTO basis (shown in dotted
  lines). Apart from the high-energy range in the last panel, where
  only V-$4s$ {\it N}MTO's are used to form the truly minimal set, the
  downfolded bands in various panels are indistinguishable from the
  bands in the full {\it N}MTO basis, within the respective energy range
  of interest. The {\it N}MTO energy points, $\epsilon _{n}$-s, spanning
  the region of interest are shown on the right-hand side in each panel.}
\label{nm_down}
\end{figure}
\end{center}
\end{widetext}

\begin{figure}
\includegraphics[width=7.2cm,keepaspectratio]{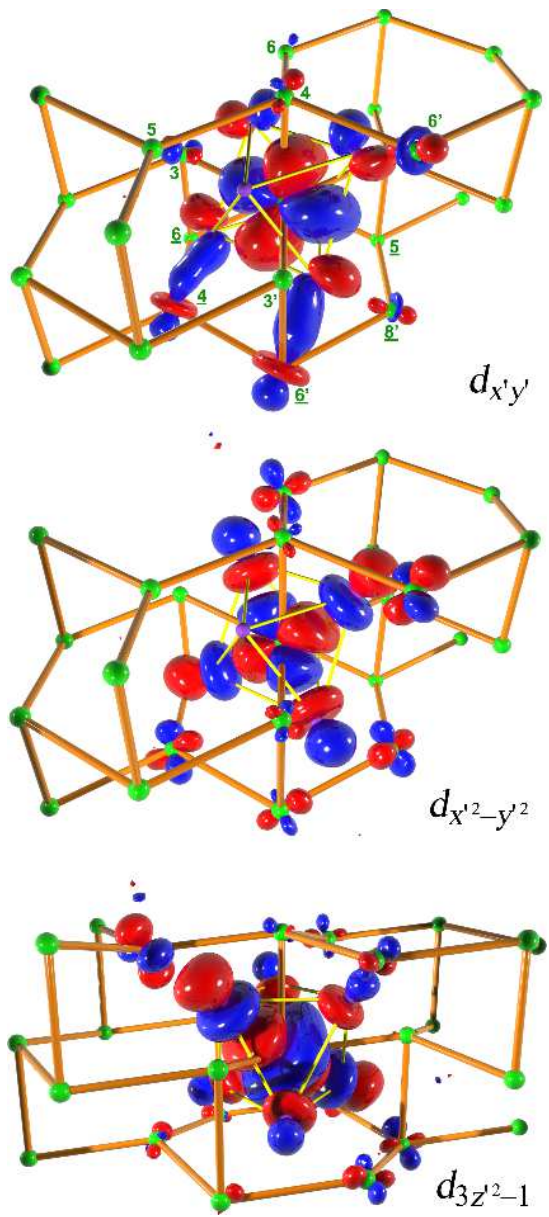}
\caption{(Color online) One of the three congruent orbitals of the truly minimal
  $t_{2g}$ {\it N}MTO set and the two orbitals of the truly minimal
  $e_{g}$ {\it N}MTO set corresponding to
  downfolded bands in FIG.\ \ref{nm_down}. Shown are the orbital shapes
  (constant-amplitude surfaces) with the lobes of opposite signs
  labeled by red and blue respectively. {\it N}MTO's are localized by
  construction: An orbital of {\it e.g.} $t_{2g}$ set is confined by
  the condition that it has no $t_{2g}$ character on other V atoms
  but may have O-$p$, V-$e_g$ characters, as is evident from the plots.}
\label{orbital1}
\end{figure}

In the first three panels of FIG.\ \ref{nm_down}, we show the bands obtained 
by using truly minimal sets, either O-$p$ or V-$t_{2g}$ or V-$e_{g}$ in solid 
lines as compared to full LDA band-structure in dotted lines. The basis sets
for the band-structure calculations shown in solid lines, which we
call as {\it downfolded} bands, contain as many orbitals as the
number of bands - hence is the name {\it truly minimal} basis set.
The {\it N}MTO-{\it downfolding} procedure enables one to construct
a set of O-$p$ or V-$t_{2g}$ or V-$e_{g}$ muffin-tin orbitals of
order {\it N}, {\it N}MTO, which span the O-$p$-like or V-$t_{2g}$-like or
V-$e_{g}$-like bands - and no other bands - with arbitrary accuracy as
{\it N} increases. Such a set is exact for the energies, 
$\epsilon _{0},....,\epsilon _{N}$, chosen for its construction.
As is seen in FIG.\ \ref{nm_down}, three energy points were used for the
construction of O-$p$, V-$t_{2g}$ and V-$e_{g}$ minimal sets, so
the MTOs are of order {\it N} = 2, {\it i.e.} they are quadratic
MTOs or QMTOs. Since the 18 O-$p$-like, 12 $t_{2g}$-like and 8
$e_g$-like bands are isolated from the above and below-lying bands,
the {\it N}MTO set obtained by making the energy mesh, $\epsilon
_{0},....,\epsilon _{N}$, finer and finer will converge to the
Hilbert space spanned by any set of Wannier functions. In other
words, the symmetrically orthogonalized set of converged {\it N}MTOs is a
set of Wannier functions. As is seen, already with choice
of three energy points, the
downfolded bands are indistinguishable from the full LDA bands in
the region of interest spanned by O-$p$, V-$t_{2g}$ and V-$e_g$
respectively - so, they are converged in the above-mentioned
sense. The corresponding {\it N}MTOs are localized by construction
as explained in the Appendix, but they are not quite
orthogonal. These truly minimal {\it N}MTO sets, therefore must be
symmetrically orthogonalized in order to become a set of localized
Wannier functions. In FIG.\ \ref{orbital1} we show one of the three congruent
orbitals of such a $t_{2g}$ {\it N}MTO set, namely $d_{x^{'}y^{'}}$ and two orbitals of the $e_g$
set before orthogonalization. Only the central part of the orbitals
have $d_{x^{'}y^{'}}$, $d_{{x^{'}}^{2}-{y^{'}}^{2}}$ or $d_{{3z^{'}}^{2}-1}$ character.
In order to describe the hybridization with the O-$p$ and
the hybridization between V-$t_{2g}$ and V-$e_g$ within the V-$d$
manifold, the O-$p$ character and, 
V-$e_g$ character for the case of $d_{x^{'}y^{'}}$ and  V-$t_{2g}$
character for the case of 
$d_{{x^{'}}^{2}-{y^{'}}^{2}}$ or $d_{{3z^{'}}^{2}-1}$, are folded into 
the tails.  In fact, {\it all} the partial wave characters {\it
other} than the respective active characters, $d_{x^{'}y^{'}}$ in the left panel,
$d_{{x^{'}}^{2}-{y^{'}}^{2}}$ in the middle panel or $d_{{3z^{'}}^{2}-1}$ in the last
panel, are folded down in the tails. 
We see the strong $pd{\sigma}$ anti-bonds in the plots 
of $e_g$ {\it N}MTOs and relatively weak $pd{\pi}$ anti-bonds in the plot 
of $t_{2g}$ {\it N}MTO.

The last panel in FIG.\ \ref{nm_down}, deals with the more difficult case, where
the chosen bands, namely the V-$s$ bands overlap with the other
high-lying bands, {\it e.g.} V-$p$, O-$d$ bands. As is seen in the
figure, even for such a difficult case of bands of interest
overlapping with other bands, it is possible to pick out the
selected bands - in the present case four V-$s$ bands arising from
four V atoms in the unit cell. With chosen two energy points, the
downfolded bands (shown in solid lines) differ from the full LDA
band-structure (shown in dotted lines). Nevertheless, we see that the bottom part of the
V-$s$ derived bands has been reproduced
quite well over an energy range of about 2 eV with merely two
energy points. As expected, increasing the number of energy points 
improves the agreement. The Wannier functions of such a complex of
bands which is overlapping with other band complexes are
ill-defined - the corresponding set of orthogonalized {\it N}MTOs, is
therefore, the set of Wannier-like functions.

As an illustrative purpose for the LMTO practitioners and to
appreciate the improvements within the {\it N}MTO procedure, in
FIG.\ \ref{lm_down} we also show the {\it downfolded} O-$p$, V-$t_{2g}$,
V-$e_{g}$ and V-$s$ computed within the framework of LMTO, using
the standard TB-LMTO code where the nearly orthogonal LMTO's are
used for producing the truly minimal basis sets. As shown in the
figure, the method works to a certain level of accuracy, provided 
all the $l$- and $R$-dependent $\epsilon_{\nu}$'s are put in the
energy region of interest and not at the center of gravity of
the respective occupied manifold as is done during the
self-consistent loops of LMTO. However, within the LMTO scheme,
the accuracy of the {\it downfolded} bands compared to the full
basis band structure is not up to the level of satisfaction and
importantly the generated minimal basis does not have the desired
Wannier-like description.

\begin{widetext}
\begin{center}
\begin{figure}
\includegraphics[width=17cm,keepaspectratio]{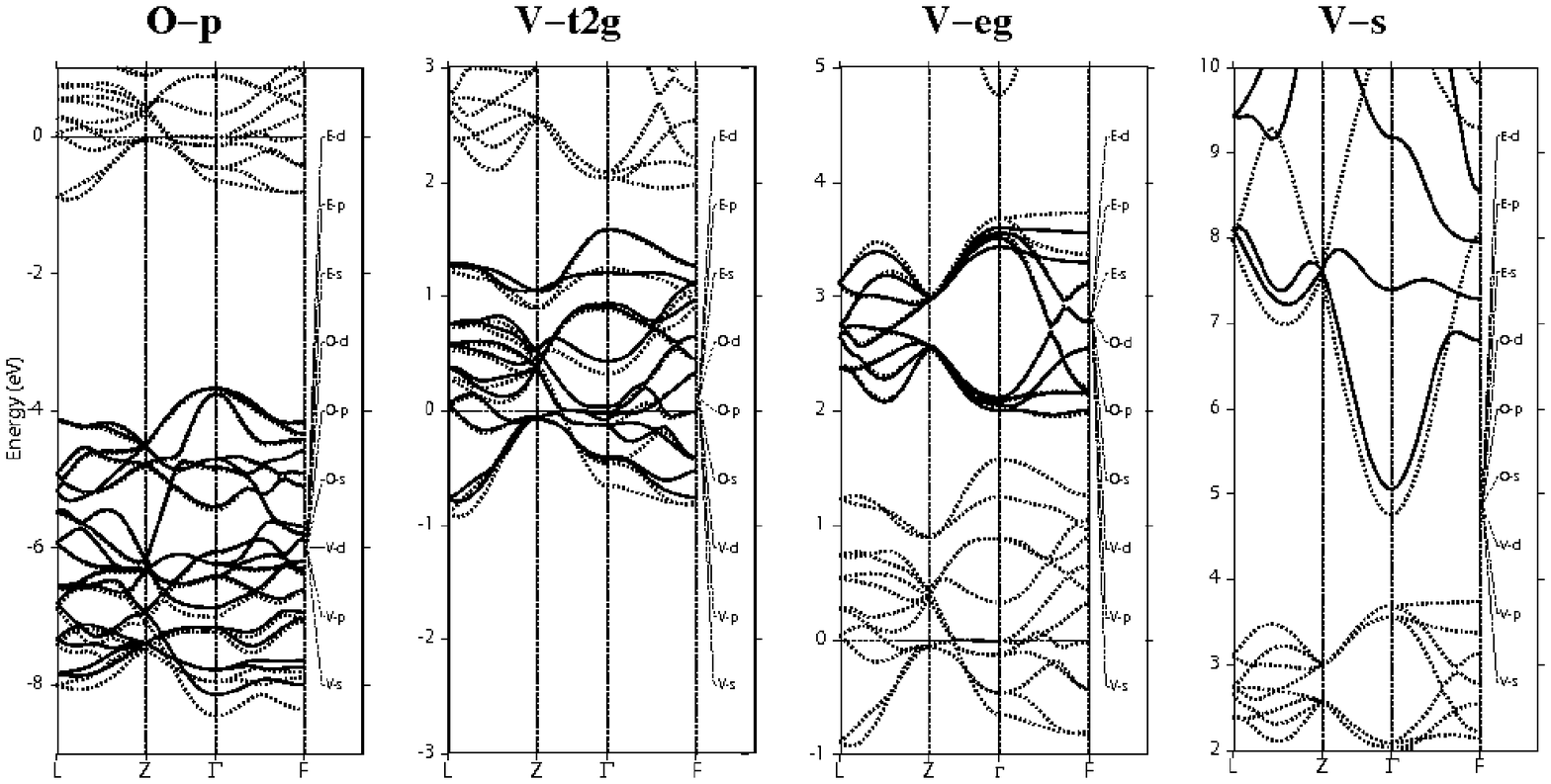}
\caption{Like in FIG.\ \ref{nm_down}, but the solid lines in different panels
  now show the bands obtained with the downfolded
  O-$2p$, V-$t2_{2g}$, V-$e_g$ and V-$4s$ LMTO basis in comparison
  to those obtained with full LMTO basis (shown in dotted
  lines). The $l$- and $R$-dependent $\epsilon _{\nu}$-s 
  are shown on the right-hand side in each panel.}
\label{lm_down}
\end{figure}
\end{center}
\end{widetext}

\subsection{Corundum V$_{2}$O$_{3}$: Tight-Binding Hamiltonian corresponding to t$_{2g}$
bands}

A reasonable approach in the tackling the V$_{2}$O$_{3}$ problem is
to start with low-energy $t_{2g}$ bands, since it is the $t_{2g}$
manifold that gets partially filled with two V electrons. In this
sub-section, we therefore zoom in on to the $t_{2g}$ bands and
discuss the tight-binding hopping integrals constructed out of the
symmetrically orthonormalized {\it N}MTOs for the truly minimal
basis set of $t_{2g}$ orbitals.

Although it would have been possible to compute the tight-binding
Hamiltonian parameters in terms of three congruent orbitals d$_{x^{'}y^{'}}$, d$_{x^{'}z^{'}}$, 
d$_{y^{'}z^{'}}$, for the sake of comparison with previous results,
we preferred to work with the orbitals of eigen representation of
symmetry lowered $D_{3d}$ group, namely the $a_{1g}$ and the two
$e_{g}^{\pi}$ orbitals which transform according to $a_{1g}$ and
$e_{g}^{\pi}$ irreducible representations. As mentioned in section
II, the unitary transformation relating $a_{1g}$ and $e_{g}^{\pi}$
with d$_{x^{'}y^{'}}$, d$_{x^{'}z^{'}}$, d$_{x^{'}z^{'}}$ is given by,

\[ \left( \begin{array}{cccc}
U & a_{1g} & e_{g}^{\pi},1  & e_{g}^{\pi},2 \\
d_{x'z'} & 1/\sqrt{3} & 0 & \sqrt{2/3} \\
d_{x'y'} & 1/\sqrt{3} & 1/\sqrt{2} & -1/\sqrt{6} \\
d_{y'z'} & 1/\sqrt{3} & -1/\sqrt{2} & -1/\sqrt{6} \\
\end{array}
\right) \]


\begin{figure}
\includegraphics[width=8.5cm,keepaspectratio]{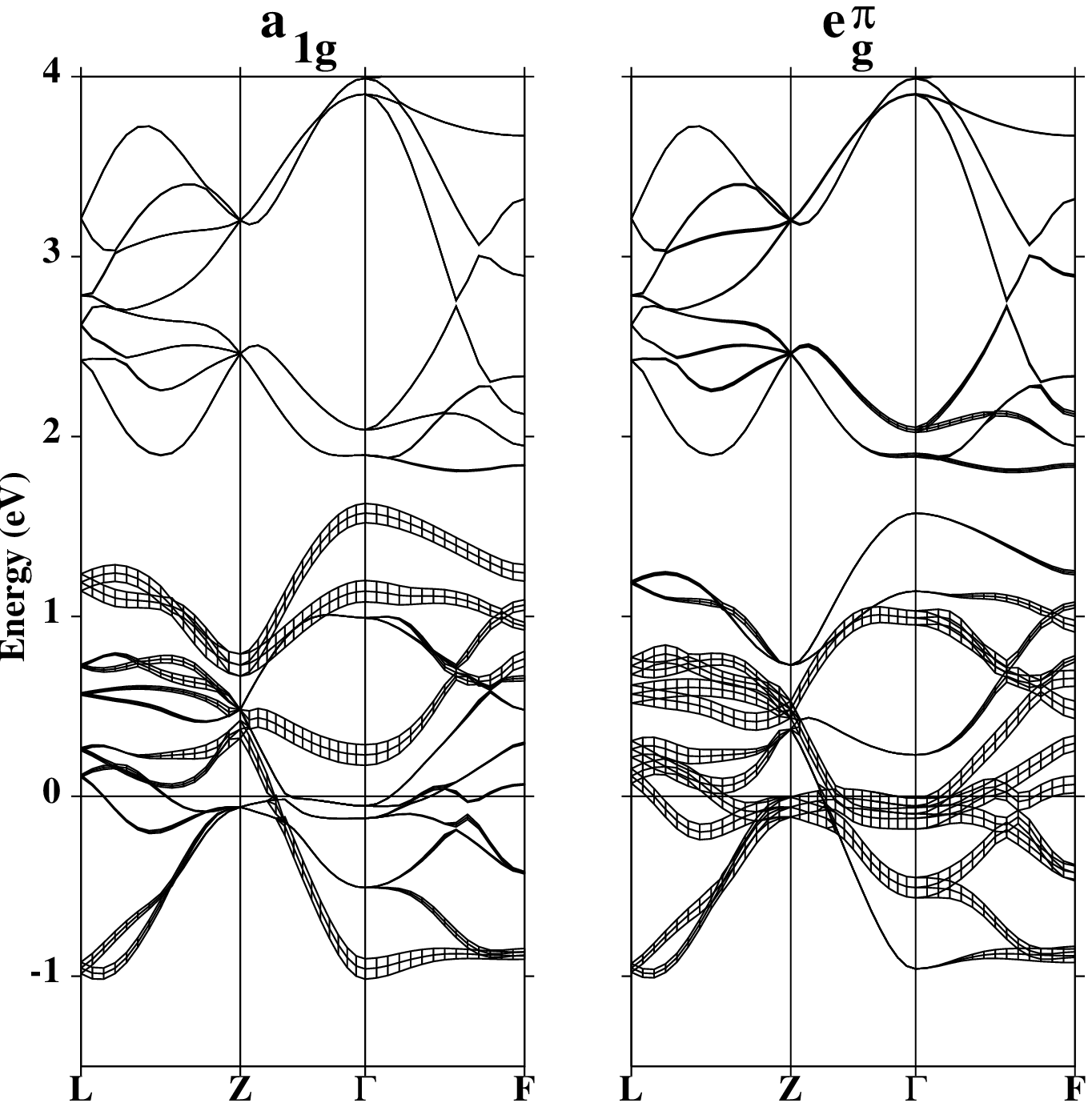}
\caption{LDA band-structure of corundum V$_{2}$O$_{3}$ with the orbital character projected on
to $a_{1g}$ and $e_{g}^{\pi}$ orbitals constructed out of three
congruent $t_{2g}$ orbitals, $x^{'}y^{'}$, $x^{'}z^{'}$ and
$y^{'}z^{'}$ [see Eqn.(1)].}
\label{fat_a1g}
\end{figure}
\begin{figure}
\includegraphics[width=8.5cm,keepaspectratio]{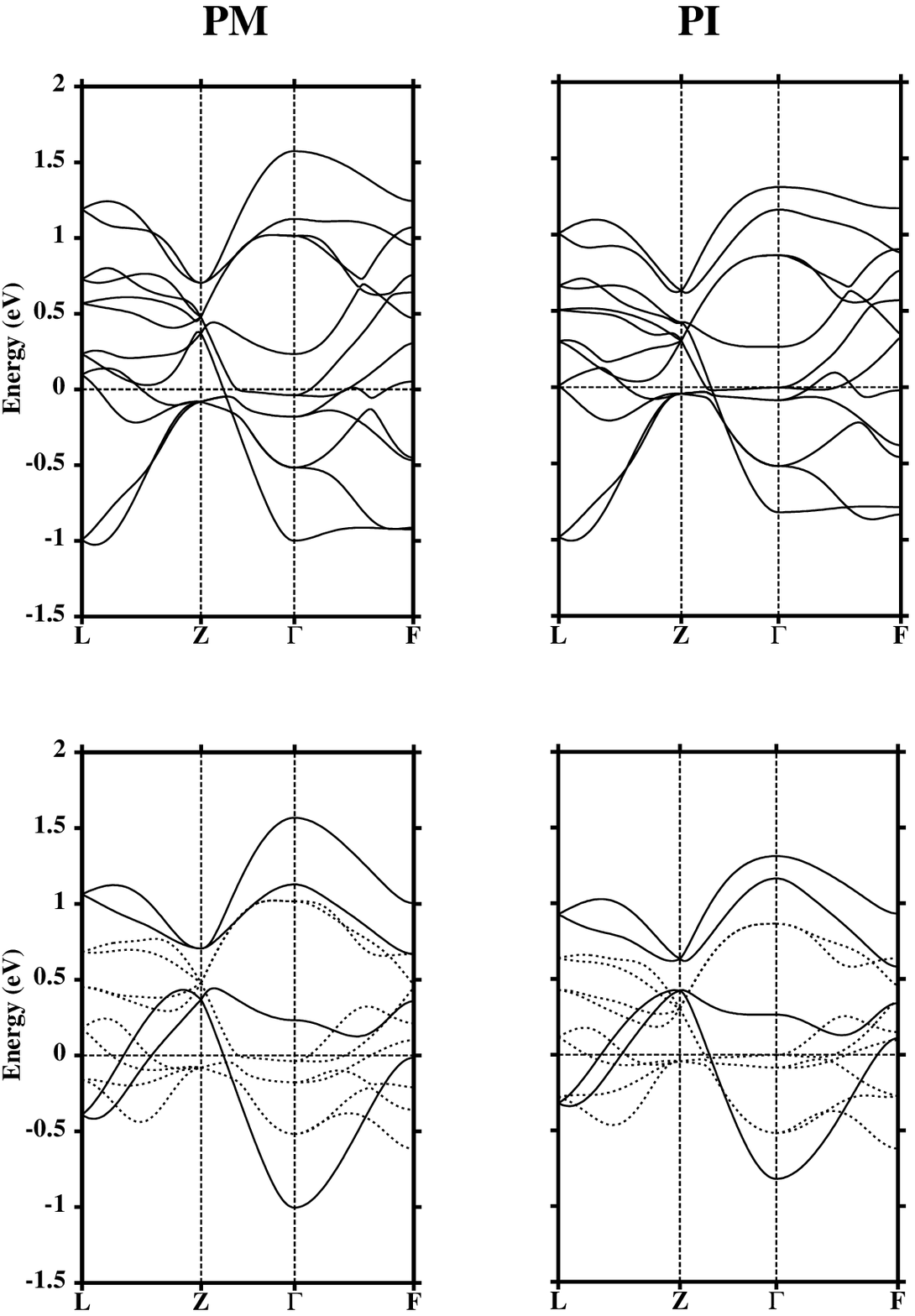}
\caption{Top panel: Downfolded band-structure of
  corundum-structured V$_{2}$O$_{3}$ in pure and Cr-doped phases obtained with the
  truly minimal basis set consisting of $a_{1g}$ and $e_{g}^{\pi}$
  {\it N}MTOs. The bands
  in the left
  are identical with the solid bands shown in the second panel of
  FIG.\ \ref{nm_down}, and that of the bands obtained with the full {\it N}MTO basis in
  the energy range -1.5 below to 2 eV above $E_f$=0. The members of the basis set are shown 
  in FIG.\ \ref{orbital2}. Bottom panel:
  The $a_{1g}$ (solid lines) and $e_{g}^{\pi}$ (dotted lines) bands
  in pure and Cr-doped V$_{2}$O$_{3}$
  where the hybridization between  $a_{1g}$ and $e_{g}^{\pi}$
  degrees of freedom has been switched off. Note the shrinkage of
  the total band-width compared to that in the left.}
\label{a1g-eg}
\end{figure}


FIG.\ \ref{fat_a1g} shows the LDA band-structure of rhombohedral V$_{2}$O$_{3}$
plotted over an energy range of -1.5 eV below the Fermi level to 4 eV above
the Fermi level, now projected on to $a_{1g}$ and $e_{g}^{\pi}$ degrees of
freedom. $e_g$ ($e_{g}^{\sigma}$) derived bands are also seen
within the energy scale of the plot. We notice significant mixing
between $a_{1g}$ and $e_{g}^{\pi}$ characters in the bands of
interest spanning the energy range -1.0 to 1.7 eV, arising due to
$a_{1g}$-$e_{g}^{\pi}$ hopping processes between neighboring V
sites. We also notice due to symmetry reason $e_{g}^{\pi}$ orbitals
acquire non-significant $e_g$ ($e_{g}^{\sigma}$) character too. 

In the top left panel of FIG.\ \ref{a1g-eg} we show the downfolded band-structure
obtained with truly minimal set consisting of $a_{1g}$ and
$e_{g}^{\pi}$. The bands, as they should be, are identical with the
downfolded bands, shown in the second panel of FIG.\ \ref{nm_down}, obtained
using d$_{x^{'}y^{'}}$, d$_{x^{'}z^{'}}$, d$_{x^{'}z^{'}}$ minimal
basis set. The bottom left of the figure shows the $a_{1g}$ and
$e_{g}^{\pi}$ bands switching off the $a_{1g}$-$e_{g}^{\pi}$
hybridization. It is important to note that the width of the
projected $a_{1g}$ band in FIG.\ \ref{fat_a1g} is much more than
that where the hybridization between $a_{1g}$ and
$e_{g}^{\pi}$ is neglected. Much of the $a_{1g}$ band width
therefore comes from the hybridization with
$e_{g}^{\pi}$.  This implies the crucial role of the $a_{1g}$ and
$e_{g}^{\pi}$ hybridization in proper description of the
band-structure.

\begin{figure}
\includegraphics[width=7.2cm,keepaspectratio]{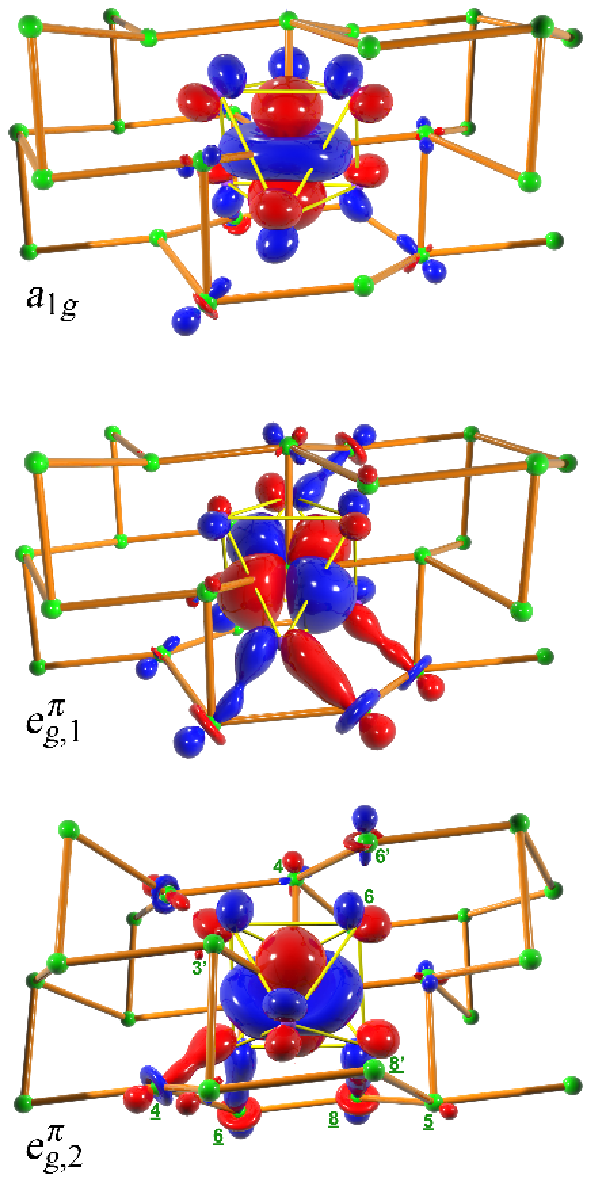}
\caption{Like in FIG.\ \ref{orbital1}, but for the $a_{1g}$ and two $e_{g}^{\pi}$
  {\it N}MTOs, corresponding to downfolded bands in FIG.\ \ref{a1g-eg}. The
  $a_{1g}$ orbital is oriented vertically while the
  $e_{g}^{\pi}$ orbitals are of more planar geometry. Due to symmetry
  reason, the $e_{g}^{\pi}$ orbitals bind strongly with $e_g$
  degrees of freedom at other V sites while this binding is
  practically negligible for $a_{1g}$.}
\label{orbital2}
\end{figure}

The members of the truly minimal set are shown in FIG.\ \ref{orbital2}, the
$a_{1g}$ orbital which is oriented vertically and two more planar
$e_{g}^{\pi}$ orbitals. Following the mixing between $e_g$
($e_{g}^{\sigma}$) and $e_{g}^{\pi}$ as seen in the {\it fatband}
plot of FIG.\ \ref{fat_a1g}, we  notice the significant presence of
$e_{g}$ tails in the plots of $e_{g}^{\pi}$
orbitals, in addition to usual anti-bonding covalent character of
the oxygen tails. The $e_{g}$ tails combine with the
oxygen tails to produce sausage-like structures {\it e.g.} that
at V sites located at $\underline{4}$ and $\underline{6}$ in
the plot of $e_{g}^{\pi},2$. 

Once we have defined the basis, in the following we compute the
tight-binding hopping matrix elements between the orthonormalized,
truly minimal $a_{1g}$ and $e_{g}^{\pi}$ {\it N}MTOs. As indicated in
the Appendix, this is done by constructing H$^{LDA}(\mathbf{k})$ in
the Bloch $\mathbf{k}$-representation, in the basis
of symmetrically orthonormalized {\it N}MTOs, $\left| \chi ^{\left(
N\right) \perp }\right\rangle$, defined for $a_{1g}$ and
$e_{g}^{\pi}$ truly minimal set, for all $\mathbf{k}$-points in the BZ and
by subsequent Fourier transformation H($\mathbf{k}$) $\longrightarrow$
H($\mathbf{r}$), 
for a given cluster with real-space range, {\it r}.
Following CNR paper, for our real-space calculation, we considered
a cluster of fourteen V sites - the V sites belonging to such a cluster
are marked in FIG.\ \ref{structure}. The convention adopted for numbering the V
atoms is same as that of CNR. All the V atoms which have at least
one  shared oxygen in the VO$_{6}$ octahedra with that of the
central V atom (marked as 1 in FIG.\ \ref{structure}) are considered in this
process.
Distance-wise the short vertical V1-V4 bond forms
the nearest-neighbor (NN), while the three more or less planar bonds
along the directions 2, 3, and 3$'$ form the 2nd neighbor shell with
distances $6 \%$ larger than the V1-V4 bond-length of 2.697
$\AA$. The next shell of neighbors {\it i.e.} the 3rd neighbor
shell is formed by the V atoms,
$\underline{4}$, $\underline{8}$ and $\underline{8^{'}}$ sitting
at a distance 0.590 $\AA$ farther than the second NN shell
consisting of V2, V3, and V3$^{'}$. The farthest shell of neighbors,
4-th neighbor shell in
the cluster are formed by the V atoms, V5, V6, V6$^{'}$,
V$\underline{5}$, V$\underline{6}$, V$\underline{6'}$ whose distances
from the central V1 atoms differs by another 6$\%$ from that of V$
\underline{4}$, V$\underline{8}$ and V$\underline{8^{'}}$.


\begin{figure}
\includegraphics[width=7.2cm,keepaspectratio]{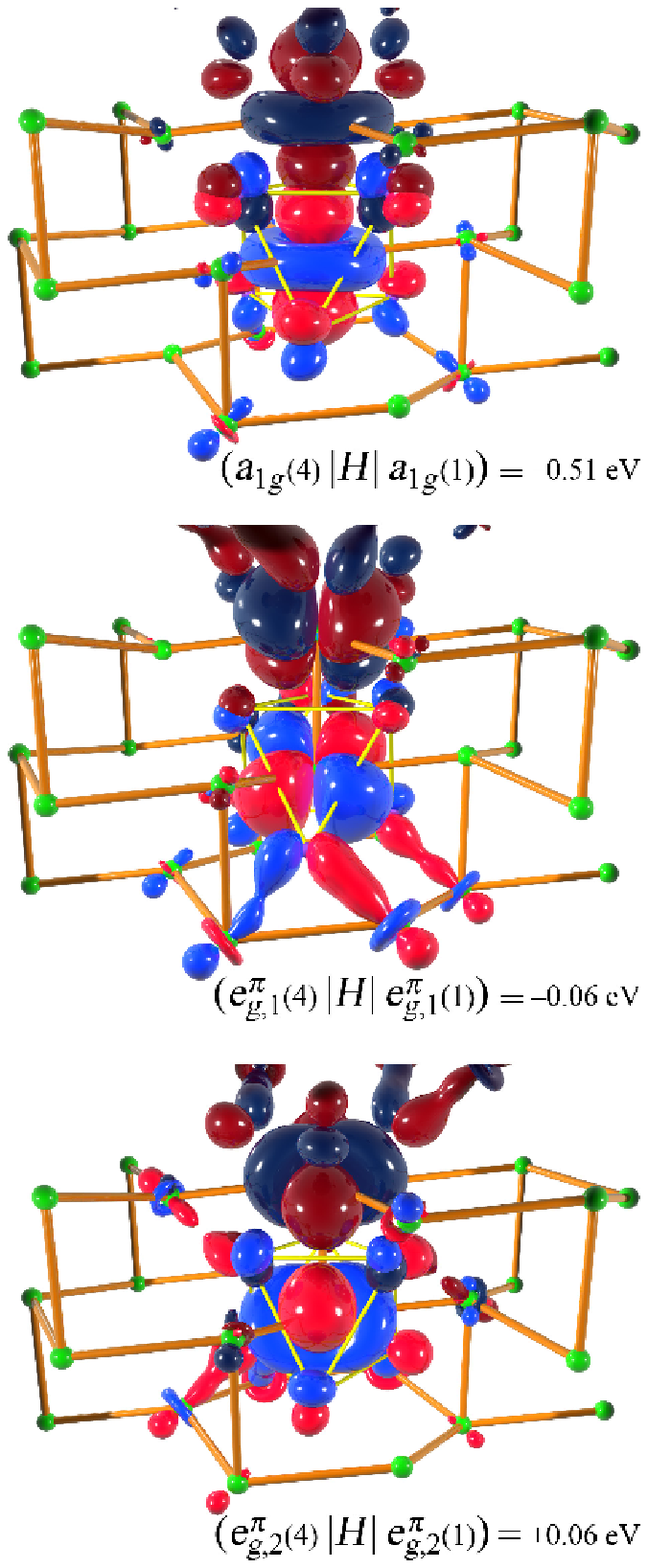}
\caption{Overlap between V-$a_{1g}$ and V-$e_g$ downfolded {\it N}MTOs,
  placed at the central V1 site and the V4 site, forming vertical
  nearest-neighbor V1-V4 pair. The light(dark) shaded orbitals
  correspond to V1(V4). This gives the idea of important
  hopping paths. Note the hopping paths via the oxygen tails in
  addition to direct V-V paths.}
\label{ovl1}
\end{figure}

\begin{figure}
\includegraphics[width=7.2cm,keepaspectratio]{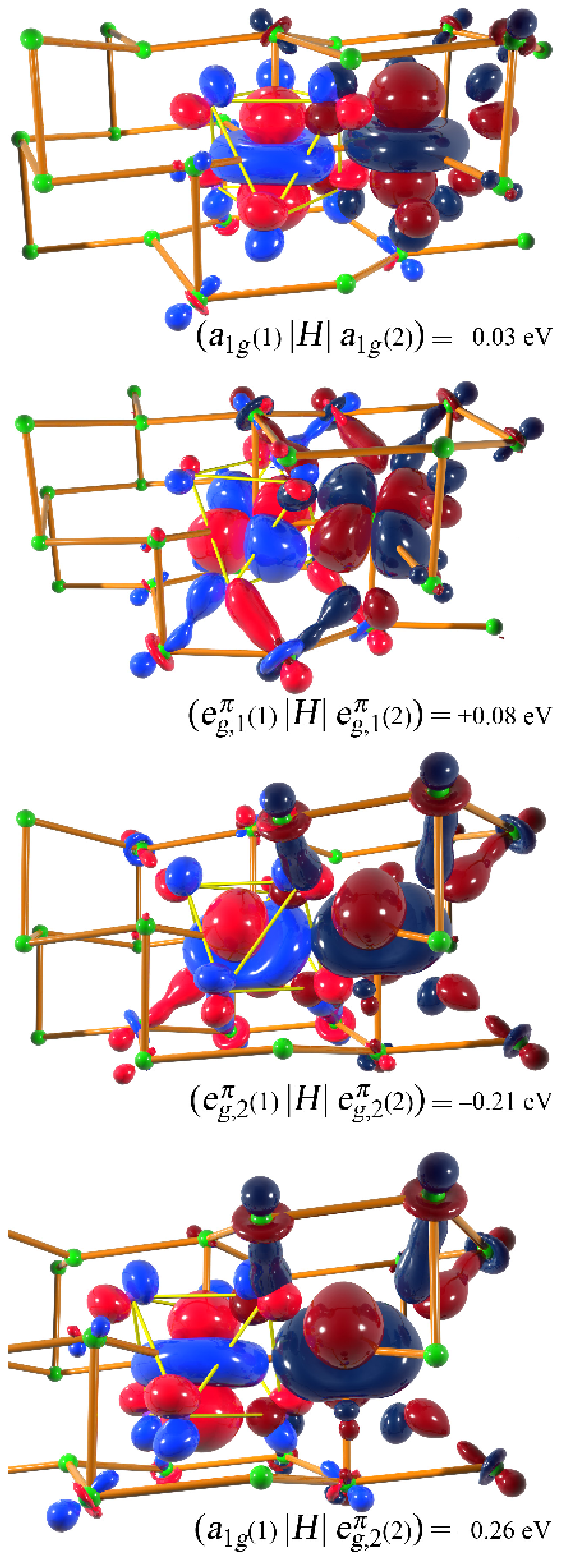}
\caption{Same as FIG.\ \ref{ovl1}, but the {\it N}MTOs are placed one at the central
V1 site and another at the neighboring V2 site along the
$x$-axis.  The light(dark) shaded orbitals
  correspond to V1(V2). Note the importance of $e_g$ tails in providing the
hopping channels, in addition to oxygen mediated and
direct V-V hopping channels.}
\label{ovl2}
\end{figure}

Before we actually attempt on numerically computing the matrix elements of
the tight-binding Hamiltonian between the orthonormalized,
downfolded $a_{1g}$ and $e_{g}^{\pi}$ {\it N}MTOs, a rough guess of the
relative strength of various hopping matrix elements may be
obtained by examining the overlap of {\it N}MTOs placed at
different V sites. In FIGs.\ \ref{ovl1} and \ref{ovl2}, we consider two such
representative cases where the truly minimal {\it N}MTOs are placed
along the vertical bond, 4, at sites V1 and V4 (FIG.\ \ref{ovl1}) and that along the
horizontal bond, 2, at sites V1 and V2 (FIG.\ \ref{ovl2}). The important
feature to notice is the hopping paths via the oxygen and the
$e_g$ tails in addition to the V-$d$ $-$ V-$d$ hopping
paths. While the approach taken by CNR, did considered the 
renormalization effect coming from oxygen degrees of freedom in
some form, the effect due to the renormalization coming from
$e_g$'s was completed ignored, which has important consequences for
hopping integrals {\it e.g.} connecting V1
and V2 sites in FIG.\ \ref{ovl2}. These additional hopping paths via the
$e_g$ tails increase the importance of the hopping processes in the
basal plane. Focusing on to the $a_{1g}-a_{1g}$ vertical pair
overlap, the biggest of all the hopping processes, we see the
large, direct  $a_{1g}-a_{1g}$ hopping which is bonding ({\it negative}) 
in nature - the red lob at V1 site overlaps with the red lob at V4
site. To understand the additional hopping contributions via the oxygen tails
let us consider the renormalized $a_{1g}$ orbitals at V1 and V4
which considering only the oxygen contributions can be written,
from a simplistic point of view as:
\begin{eqnarray}
\vert \Psi_{1} \rangle & = & \vert d_{1} \rangle + \lambda \vert
p_{1} \rangle \\ \nonumber
\vert \Psi_{4} \rangle & = & \vert d_{4} \rangle + \lambda \vert
p_{4} \rangle
\label{covalen} 
\end{eqnarray}
where $\lambda$ is the covalency mixing parameter between
V-$a_{1g}$ and O-$p$, $\vert p_{1} \rangle$ and  $\vert p_{4} \rangle$
are the wave-function of the shared O's between the V1O$_{6}$
and V4O$_{6}$ octahedra having the same $a_{1g}$ symmetry,$\vert
d_{1} \rangle$ and  $\vert d_{4} \rangle$ are the bare $a_{1g}$
orbitals at the V1 and V4 sites. From this, we see that the overlap
between the renormalized $a_{1g}$ orbitals at sites V1
and V4 is given as , 
\[
\langle  \Psi_{1}, \Psi_{4} \rangle = \langle  d_{1}, d_{4}
\rangle + \lambda  ( \langle  p_{1}, d_{4}
\rangle +  \langle  d_{1}, p_{4} \rangle ) + \lambda^{2} \langle  p_{1}, p_{4}
\rangle
\]
While both the $d-d$ and $p-p$ overlaps, $\langle  d_{1}, d_{4}
\rangle$ and $\langle  p_{1}, p_{4} \rangle$ are of bonding
nature (the contribution $\langle  p_{1}, p_{4}
\rangle$ is small due to the presence of prefactor $\lambda^{2}$),
the sign of the correction terms, $\langle  p_{1}, d_{4}
\rangle$ and $\langle  d_{1}, p_{4} \rangle$ depend on the
V1-O-V4 angle, which for the real crystal turn out to be
82.3$^{o}$. Careful investigation of FIG.\ \ref{ovl1} show these overlaps
to be anti-bonding ({\it positive})\cite{note_orb}.  These
anti-bonding hopping paths, therefore, oppose the bonding, direct
$a_{1g}-a_{1g}$ hopping, and thereby reduces the magnitude of the
effective $a_{1g}-a_{1g}$ hopping from the bare $a_{1g}-a_{1g}$
hopping. We will return to this point again while discussing the
crucial sensitivity of this important hopping parameter on the
intricate details of the geometrical structure. We further see that
the direct vertical hopping processes between $e_g^{\pi}$s are
rather weak, which get weaker by the oxygen renormalization effect.

Moving to FIG.\ \ref{ovl2}, for overlap along the horizontal bond direction,
2, we see a weak overlap between the $a_{1g}$ truly minimal {\it N}MTOs
while the $e_g$ like tails make the hopping between $e_{g}^{\pi},2$
and $e_{g}^{\pi},2$ nearly as strong as that between  $a_{1g}$
and $e_{g}^{\pi},2$. The overlap $e_{g}^{\pi},1 - e_{g}^{\pi},1$ is
anti-bonding ({\it positive}) while $e_{g}^{\pi},2 - e_{g}^{\pi},2$
and $a_{1g} - e_{g}^{\pi},2$
overlaps are bonding ({\it negative}).

\begin{table}
\begin{tabular}{rrrrrrrrrrr} \hline
&  &  &  &  & $\epsilon$ & (eV) &  &  &  &  \\ 
&  &  &  &  &  &  &  &  &  &  \\ 
& PM & $a_{1g}$ & $e_{g,1}^{\pi }$ & $e_{g,2}^{\pi }$ & $e_{g,1}^{\pi }$ & $%
e_{g,2}^{\pi }$ & $a_{1g}$ & $e_{g,1}^{\pi }$ & $e_{g,2}^{\pi }$ & $a_{1g}$
\\ 
& PI & $\downarrow $ & $\downarrow $ & $\downarrow $ & $\downarrow $ & $%
\downarrow $ & $\downarrow $ & $\downarrow $ & $\downarrow $ & $\downarrow $
\\ 
& AFI & $a_{1g}$ & $e_{g,1}^{\pi }$ & $e_{g,2}^{\pi }$ & $e_{g,2}^{\pi }$ & $%
e_{g,1}^{\pi }$ & $e_{g,1}^{\pi }$ & $a_{1g}$ & $a_{1g}$ & $e_{g,2}^{\pi }$
\\ 
&  &  &  &  &  &  &  &  &  &  \\ 
$1$ &  & $.27$ & $.00$ & $.00$ & $.00$ & $.00$ & $.00$ & $.00$ & $.00$
& $.00$ \\ 
$\downarrow $ &  & $.30$ & $.00$ & $.00$ & $.00$ & $.00$ & $.00$ & $.00$
& $.00$ & $.00$ \\ 
$1$ &  & $.28$ & $.00$ & $.00$ & $.01$ & $.01$ & $-.04$ & $-.04$ & $.00$
& $.00$ \\ \hline
\end{tabular}
\caption{On-site matrix elements in the high-temperature paramagnetic metallic phase
(undoped, ambient pressure V$_{2}$O$_{3}$),
paramagnetic insulating ((V$_{0.962}$Cr$_{0.038}$)$_{2}$O$_{3}$) and
in the low-temperature antiferromagnetic insulating (monoclinic) phase, between the $m$-orbital 
  and the $m'$-orbital. Except for the orthonormalization, the
  orbitals are as defined in FIG.\ \ref{orbital2}. 
        }
\end{table}

\begin{table}
\begin{tabular}{rrrrrrrrrrr}\hline
$d$ & ($\AA$) &  &  &  & $t$ & (eV) &  &  &  &  \\ 
& PM & $a_{1g}$ & $e_{g,1}^{\pi }$ & $e_{g,2}^{\pi }$ & $e_{g,1}^{\pi }$ & $%
e_{g,2}^{\pi }$ & $a_{1g}$ & $e_{g,1}^{\pi }$ & $e_{g,2}^{\pi }$ & $a_{1g}$
\\ 
& PI & $\downarrow $ & $\downarrow $ & $\downarrow $ & $\downarrow $ & $%
\downarrow $ & $\downarrow $ & $\downarrow $ & $\downarrow $ & $\downarrow $
\\ 
& AFI & $a_{1g}$ & $e_{g,1}^{\pi }$ & $e_{g,2}^{\pi }$ & $e_{g,2}^{\pi }$ & $%
e_{g,1}^{\pi }$ & $e_{g,1}^{\pi }$ & $a_{1g}$ & $a_{1g}$ & $e_{g,2}^{\pi }$
\\ 
&  &  &  &  &  &  &  &  &  &  \\ 
$1$ & 2.70 & $-.51$ & $-.06$ & $.06$ & $.00$ & $.00$ & $.00$ & $.00$ & $.00$
& $.00$ \\ 
$\downarrow $ & 2.75 & $-.43$ & $-.07$ & $.07$ & $.00$ & $.00$ & $.00$ & $.00$
& $.00$ & $.00$ \\ 
$4$ & 2.77 & $-.44$ & $-.06$ & $.06$ & $.00$ & $.00$ & $.00$ & $.00$ & $.00$
& $.00$ \\ 
&  &  &  &  &  &  &  &  &  &  \\ 
$1$ & 2,88 & $-.03$ & $.08$ & $-.21$ & $.00$ & $.00$ & $.00$ & $.00$ & $-.26$
& $-.26$ \\ 
$\downarrow $ & 292 & $-.02$ & $.07$ & $-.19$ & $.00$ & $.00$ & $.00$ & $.00$
& $-.24$ & $-.24$ \\ 
$2$ & 2.99 & $.02$ & $.04$ & $-.14$ & $.00$ & $.00$ & $.00$ & $.00$ & $-.19$
& $-.19$ \\ 
&  &  &  &  &  &  &  &  &  &  \\ 
$1$ & 2.88 & $-.03$ & $-.14$ & $.01$ & $-.13$ & $-.13$ & $.23$ & $.23$ & $.13$
& $.13$ \\ 
$\downarrow $ & 2.92 & $-.02$ & $-.13$ & $.01$ & $-.11$ & $-.11$ & $.21$ & $%
.21$ & $.12$ & $.12$ \\ 
$3^{\prime }$ & 2.88 & $-.03$ & $-.14$ & $.00$ & $-.14$ & $-.14$ & $.25$ & $%
.25$ & $.12$ & $.12$ \\ 
&  &  &  &  &  &  &  &  &  &  \\ 
$1$ & 2.88 & $-.03$ & $-.14$ & $.01$ & $.13$ & $.13$ & $-.23$ & $-.23$ & $.13$
& $.13$ \\ 
$\downarrow $ & 2.92 & $-.02$ & $-.13$ & $.01$ & $.11$ & $.11$ & $-.21$ & $%
-.21$ & $.12$ & $.12$ \\ 
$3$ & 2.86 & $-.07$ & $-.15$ & $.02$ & $.14$ & $.14$ & $-.25$ & $-.25$ & $.14$
& $.14$ \\ 
&  &  &  &  &  &  &  &  &  &  \\ 
$1$ & 3.47 & $-.12$ & $-.09$ & $-.04$ & $-.04$ & $-.04$ & $-.09$ & $-.09$ & $%
-.05$ & $-.05$ \\ 
$\downarrow $ & 3.45 & $-.12$ & $-.11$ & $-.05$ & $-.02$ & $-.02$ & $-.10$ & $%
-.10$ & $-.03$ & $-.03$ \\ 
$\underline{4}$ & 3.44 & $-.13$ & $-.09$ & $-.06$ & $-.03$ & $-.03$ & $-.10$
& $-.10$ & $-.05$ & $-.05$ \\ 
&  &  &  &  &  &  &  &  &  &  \\ 
$1$ & 3.47 & $-.12$ & $.01$ & $.06$ & $.02$ & $-.09$ & $.09$ & $.00$ & $.10$
& $-.06$ \\ 
$\downarrow $ & 3.45 & $-.12$ & $.01$ & $.07$ & $.05$ & $-.09$ & $.08$ & $.02$
& $.10$ & $-.07$ \\ 
$\underline{8^{\prime }}$ & 3.46 & $-.12$ & $.00$ & $.06$ & $.04$ & $-.10$ & $%
.10$ & $.03$ & $.09$ & $-.06$ \\ 
&  &  &  &  &  &  &  &  &  &  \\ 
$1$ & 3.47 & $-.12$ & $.01$ & $.06$ & $-.09$ & $.02$ & $.00$ & $.09$ & $-.06$
& $.10$ \\ 
$\downarrow $ & 3.45 & $-.12$ & $.01$ & $.07$ & $-.09$ & $.05$ & $.02$ & $.08$
& $-.07$ & $.10$ \\ 
$\underline{8}$ & 346 & $-.12$ & $.00$ & $.06$ & $-.10$ & $.04$ & $.03$ & $%
.10$ & $-.06$ & $.09$ \\ 
&  &  &  &  &  &  &  &  &  &  \\ 
$1$ & 3.69 & $-.06$ & $.01$ & $.00$ & $.09$ & $.02$ & $.00$ & $.04$ & $.00$ & 
$-.10$ \\ 
$\downarrow $ & 3.70 & $-.05$ & $.01$ & $.00$ & $.09$ & $.01$ & $.00$ & $.04$
& $.00$ & $-.12$ \\ 
$5$ & 3.63 & $-.06$ & $.00$ & $.00$ & $.10$ & $.00$ & $.00$ & $.04$ & $.03$ & 
$-.15$ \\ 
&  &  &  &  &  &  &  &  &  &  \\ 
$1$ & 3.69 & $-.06$ & $.01$ & $.00$ & $.02$ & $.09$ & $.04$ & $.00$ & $-.10$
& $.00$ \\ 
$\downarrow $ & 3.70 & $-.05$ & $.01$ & $.00$ & $.01$ & $.09$ & $.04$ & $.00$
& $-.10$ & $.00$ \\ 
$\underline{5}$ & 3.63 & $-.05$ & $.00$ & $.00$ & $.00$ & $.10$ & $.04$ & $.00
$ & $-.15$ & $.03$ \\ 
&  &  &  &  &  &  &  &  &  &  \\ 
$1$ & 3.69 & $-.06$ & $.03$ & $.02$ & $.04$ & $.08$ & $.10$ & $-.03$ & $-.03$
& $.07$ \\ 
$\downarrow $ & 3.70 & $-.05$ & $.04$ & $.03$ & $.02$ & $.08$ & $.10$ & $-.01$
& $-.04$ & $.06$ \\ 
$6^{\prime }$ & 3.74 & $-.05$ & $.04$ & $.02$ & $.02$ & $.05$ & $.09$ & $.00$
& $-.03$ & $.06$ \\ 
&  &  &  &  &  &  &  &  &  &  \\ 
$1$ & 3.69 & $-.06$ & $.03$ & $.02$ & $.08$ & $.04$ & $-.03$ & $.10$ & $.07$
& $-.03$ \\ 
$\downarrow $ &  & $-.05$ & $.04$ & $.03$ & $.08$ & $.02$ & $-.01$ & $.10$ & 
$.06$ & $-.04$ \\ 
$\underline{6^{\prime }}$ & 3.70 & $-.05$ & $.04$ & $.02$ & $.05$ & $.02$ & $%
.00$ & $.09$ & $.06$ & $-.03$ \\ 
&  &  &  &  &  &  &  &  &  &  \\ 
$1$ & 3.69 & $-.06$ & $-.03$ & $-.04$ & $.04$ & $.07$ & $-.10$ & $-.01$ & $.04
$ & $.06$ \\ 
$\downarrow $ & 3.70 & $-.05$ & $-.03$ & $-.05$ & $.04$ & $.07$ & $-.10$ & $%
-.03$ & $.03$ & $.07$ \\ 
$6$ & 3.73 & $-.07$ & $-.02$ & $-.04$ & $.07$ & $.07$ & $-.11$ & $.00$ & $.04$
& $.06$ \\ 
&  &  &  &  &  &  &  &  &  &  \\ 
$1$ & 3.69 & $-.06$ & $-.03$ & $-.04$ & $.07$ & $.04$ & $-.01$ & $-.10$ & $.06
$ & $.04$ \\ 
$\downarrow $ & 3.70 & $-.05$ & $-.03$ & $-.04$ & $.07$ & $.04$ & $-.03$ & $%
-.10$ & $.07$ & $.03$ \\ 
$\underline{6}$ & 3.73 & $-.07$ & $-.02$ & $-.04$ & $.07$ & $.07$ & $.00$ & $%
-.11$ & $.06$ & $.04$ \\ \hline
\end{tabular}
\caption{Hopping integrals in the high-temperature paramagnetic metallic,
paramagnetic insulating and
low-temperature antiferromagnetic insulating phase, from the
  central V1 site of the cluster to the Vn
  site of the cluster, where n = 4, 2, 3, 3$^{'}$, $\underline{4}$,
  $\underline{8}$, $\underline{8^{'}}$, 5, 6, 6$^{'}$,
  $\underline{5}$, $\underline{6}$ and
  $\underline{6^{'}}$.
        }
\end{table}

In the first row of blocks in Table II we show all the hopping integrals between the
central V atom (1) and the neighboring V atoms (for numbering see FIG. 1) up to 4-th
neighbor obtained 
by Fourier transform of the downfolded $t_{2g}$ Hamiltonian in 
$a_{1g}-e_{g}^{\pi}$ basis in symmetrically orthogonalized
representation. The on-site energies are shown in Table I.
We see that the hoppings beyond the four predominant directions, 4, 2, 3, 3$^{'}$ also have
non-negligible contributions. The tight-binding $t_{2g}$ 
bands considering interactions till 2NN, 3NN and 4NN hoppings are 
shown in FIG.\ \ref{tbbands}. The $k$-space band-structure considering the
infinite summation in the real-space Fourier series is also
shown for comparison. As we see, hoppings till 4NN are essential
to reproduce atleast the gross features of the band-structure.

The symmetry properties of the corundum structure allows one
to recast the important hopping integrals in the directions 4, 2, 3
and 3$^{'}$ in terms of reduced parameters
like $\mu$, $\lambda$, $\alpha$, $\beta$, $\sigma$ and $\tau$. The 
relationship of the various hopping integrals and the reduced
parameters are shown in Table III \cite{note_sign}.

\begin{table}
\begin{tabular}{ccccc} \hline
Direction & 4 & 2 & 3${'}$ & 3 \\
 &  & & &   \\ \hline
$e_{g}^{\pi }1,e_{g}^{\pi }1$ & -$\mu$ & -$\alpha$ & -1/4 $\alpha$ + 3/4 $\beta$ & -1/4
 $\alpha$ + 3/4 $\beta$  \\
$e_{g}^{\pi }2,e_{g}^{\pi }2$ &  $\mu$  \enskip \enskip & $\beta$
 \enskip \enskip & -3/4 $\alpha$ + 1/4 $\beta$ \enskip \enskip & -3/4 $\alpha$ + 1/4 $\beta$ \\
$a_{1g},a_{1g}$ & $\rho$ & $\sigma$ & $\sigma$ & $\sigma$ \\
$e_{g}^{\pi }1,e_{g}^{\pi }2$ &  0 & 0 & $\sqrt{3}/4$ ($\alpha$ + $\beta$) &  -$\sqrt{3}/4$($\alpha$ + $\beta$) \\  
$e_{g}^{\pi }1,a_{1g}$ & 0 & 0 & $\sqrt{3}/2$ $\tau$ & - $\sqrt{3}/2$ $\tau$\\
$e_{g}^{\pi }2,a_{1g}$ & 0 & -$\tau$ & 1/2 $\tau$ & 1/2 $\tau$   \\ 
$e_{g}^{\pi }2,e_{g}^{\pi }1$  &  0 & 0 & $\sqrt{3}/4$ ($\alpha$ + $\beta$) &  -$\sqrt{3}/4$($\alpha$ + $\beta$) \\ 
$a_{1g},e_{g}^{\pi }1$   &  0 & 0 & $\sqrt{3}/2$ $\tau$ & - $\sqrt{3}/2$ $\tau$\\
$a_{1g},e_{g}^{\pi }2$ &  0 & -$\tau$ & 1/2 $\tau$ & 1/2 $\tau$ \\ \hline
\end{tabular}
\caption{ Hopping parameters along the directions 4,2, 3 and 3$^{'}$ in
        terms of corundum-symmetry-adopted reduced parameters $\mu$, $\rho$, $\alpha$,
        $\beta$, $\sigma$ and $\tau$.
        }
\end{table}

In CNR's  original
paper\cite{cast} as well as in the later paper by Di Matteo {\it
et. al.} \cite{matteo}, the hopping integrals have been quoted in 
terms of these parameters. Di Matteo {\it et. al.}\cite{matteo} quoted the
estimate obtained by CNR as well as the estimates obtained by
TB fitting performed on LAPW band calculation of
Mattheiss \cite{matt}. These two estimates were found to be numerically
not very different, although it was not clear whether the tight-binding fitting
procedure included the correction due to the renormalization
effect from the oxygen degrees of freedom or not. The estimate for
the vertical pair hopping quoted by Di Matteo {\it
et. al.}\cite{matteo} appeared to be virtually same as the
estimate of $dd\sigma$ obtained by Mattheiss\cite{matt}, although one expects
some difference due to integrated out oxygen degrees of freedom.
For the sake of direct comparison and for the sake of future analysis,
we have extracted these reduced parameters from the estimate of our
hopping integrals. In Table IV we show the reduced 
parameters obtained by {\it N}MTO-{\it downfolding} technique in comparison to that of CNR
and Di Matteo {\it et. al.} for the directions 4, 2, 3$^{'}$ and 3.

\begin{widetext}
\begin{center}
\begin{figure}
\includegraphics[width=17cm,keepaspectratio]{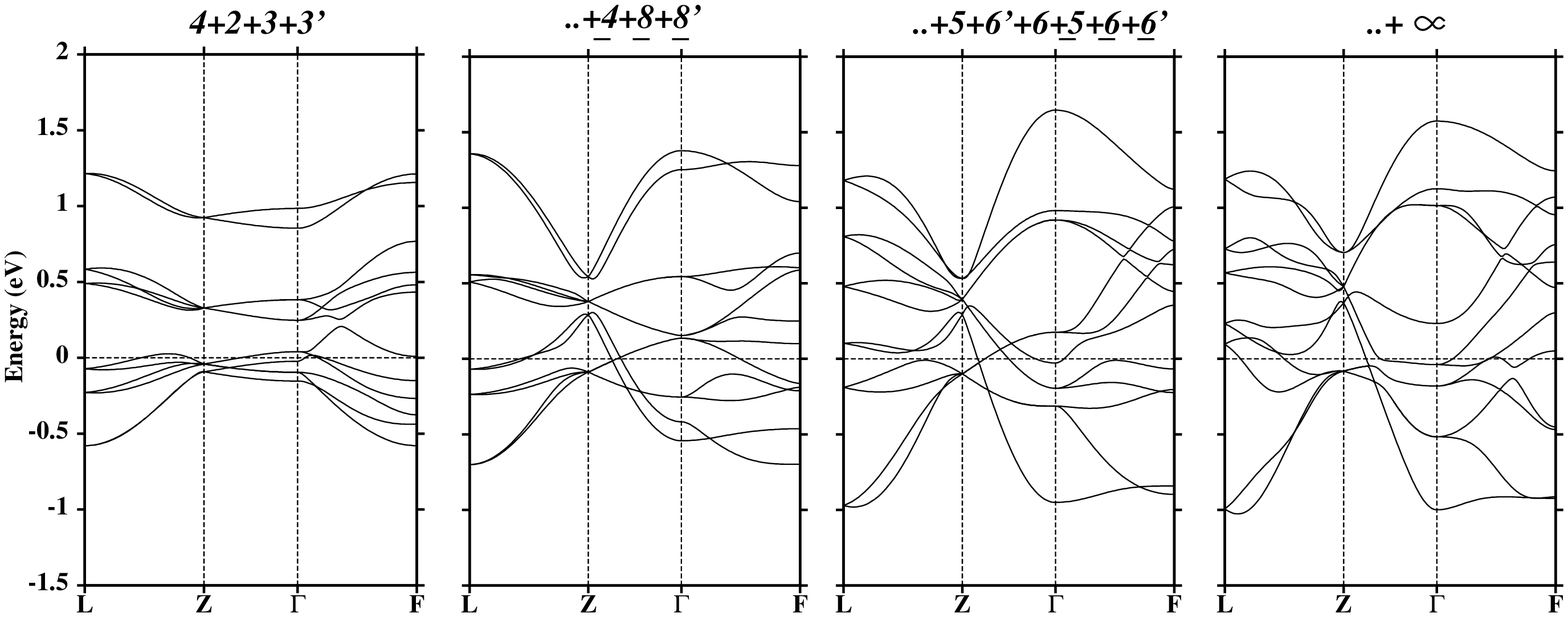}
\caption{The tight-binding bands of corundum-structured
  V$_{2}$O$_{3}$ with various range of hopping
  interactions. In the first panel, hopping interactions include only the 
  near neighbor interactions, interaction along the short vertical
  bond direction 4 and that along three basal bond directions, 2,3
  and 3$^{'}$ . In the second panel interactions until
  third nearest neighbors (directions 4, 2, 3, 3$^{'}$, $\underline{4}$,
  $\underline{8}$ and $\underline{8^{'}}$) and in the third panel interactions until
  forth nearest neighbors (directions 4, 2, 3, 3$^{'}$, $\underline{4}$,
  $\underline{8}$, $\underline{8^{'}}$, 5, 6, 6$^{'}$,
  $\underline{5}$, $\underline{6}$ and
  $\underline{6^{'}}$) are included. The last panel shows the fully converged  tight-binding 
  band-structure involving all the hopping interactions ranging
  from nearest-neighbor to infinity.}
\label{tbbands}
\end{figure}
\end{center}
\end{widetext}

\begin{table}
\begin{tabular}{cccc} \hline
 & CNR & Di Matteo et. al. & N-MTO  \\ 
 &  &  &   \\ \hline
$\mu$ & .20 & .20 & .06  \\
$\rho$ & -.72 & -.82 & -.51  \\
- $\alpha$ & -.13 & -.14 & .08 \\
$\beta$ & -.04 & -.05 & -.21  \\
$\sigma$ & .05 & .05 & -.03  \\
- $\tau$ & -.23 & -.27 & -.26  \\\hline
\end{tabular}
\caption{Comparison of TB parameters of corundum-structured
 V$_{2}$O$_{3}$ (in terms reduced parameters
 $\mu$, $\rho$, $\alpha$, $\beta$, $\sigma$ and $\tau$) obtained by
 different procedures for V-V hopping along the four
 near directions, $4$, $2$, $3^{'}$ and $3$.}
\end{table}

On examining the {\it N}MTO derived $\mu$, $\lambda$, $\alpha$, $\beta$,
$\sigma$ and $\tau$  parameters in comparison to that of 
CNR and Di Matteo {\it et. al.} we find that the parameters are quite 
different from their estimates. In particular we notice the
significant reduction of the vertical pair $a_{1g}-a_{1g}$ hopping,
given by the parameter $\rho$ and the increased importance of the
hoppings in the basal plane. In order to investigate the probable
reasons for such discrepancy, one of the prime candidate on first
glance appears to be the sophisticated treatment of {\it N}MTO-downfolding
over that of CNR, where the effective orbitals were constructed
following Anderson's super-exchange idea, the covalency V-$d-$O-$p$
mixing parameter, $\lambda$ and the charge transfer gap $E_{3d} -
E_{2p}$ were extracted from nuclear magnetic resonance and photo-emission experimental
measurements which were known only to certain accuracy, and finally
the matrix elements were computed in terms of second order
perturbation theory in $\lambda$. Nevertheless, in spite of all
the above mentioned approximations - some of which are crude- it
turns out that the structural information plays a even more crucial
role.

We consider in the following the specific case of vertical $a_{1g}-a_{1g}$
hopping which doesn't have the additional complexity of hopping via
the $e_g$ tails, another crucial ingredient not taken into account
in CNR's study. In CNR paper, the direct $d-d$ hoppings were obtained from
a linear combination of atomic orbital (LCAO) kind of approach by Ashkenazi {\it et. a.} \cite{ashkenazi} which
assumes the correct geometry, while the trigonal distortion was
assumed to negligible (set to zero)\cite{matteo}. Such an
approximation is found to have a deeper implication in terms of the
quantitative estimates of the effective $V-V$ hopping. In order to
have an understanding of the delicate effect of the geometry, we
carried out calculations on crystal structures with varying
trigonal distortions. Crystal structures with varying amount of
trigonal distortions are generated by linear interpolation of the
internal parameters associated with V and O atoms between that of
the real crystal and that of the ideal hexagonal arrangement:
\begin{eqnarray*}
z_V & = & 0.3333 (1-c) + 0.34630 c \\
x_O & = & 0.3333 (1-c) + 0.31164 c 
\end{eqnarray*}

\begin{figure}
\rotatebox{0}{\includegraphics[width=10cm,keepaspectratio]{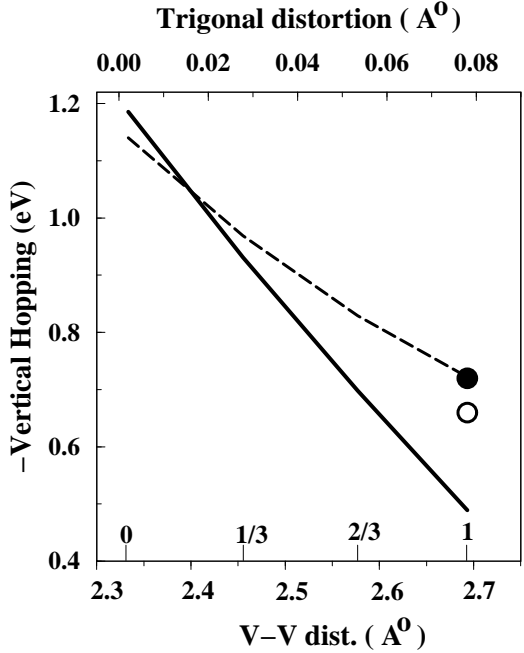}}
\caption{The influence of the trigonal distortion and V1-V4
  distance on the vertical $a_{1g}-a_{1g}$ pair hopping magnitude. Structures
  with varying amount of distortions are generated by linear
  interpolation of the internal parameters associated with V and O
  positions, $z_v = 0.3333 (1-c) + 0.34630 c$; $z_O = 0.3333 (1-c)
  + 0.311640 c$. This leads to simultaneous variation of the V1-V4
  distance and the trigonal distortion. The dotted and the solid lines
  give the estimate of the direct $a_{1g}-a_{1g}$ hopping and
  that of the effective  $a_{1g}-a_{1g}$ hopping in truly minimal
  $a_{1g}-e_{g}^{\pi}$ {\it N}MTO basis. The open and solid
  circles are the estimates of direct and effective
  $a_{1g}-a_{1g}$ hopping respectively, as obtained by
  CNR\cite{cast}. The V1-V4 distance in CNR's calculation were
  considered to be same as that in real
  structure while the amount of trigonal distortion was set to zero.}
\label{distortion}
\end{figure}

\begin{figure}
\includegraphics[width=9cm,keepaspectratio]{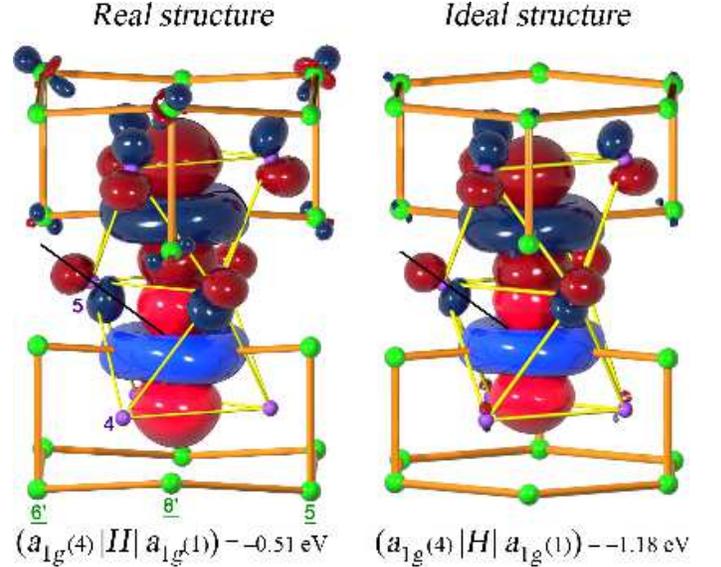}
\caption{Comparison between V1-$a_{1g}$ and V4-$a_{1g}$  overlaps
  for the real and ideal structures. The light(dark) shaded orbitals
  correspond to V1(V2). The tails of the orbital at V1 site has been
  omitted for clarity.}
\label{ovl3}
\end{figure}

Putting $c$ = 0(1) gives the ideal(real) structure. Increasing $c$
increases the trigonal distortion which is
the difference between two sets of V-O distances in VO$_{6}$
octahedra, with three short and
three long V-O bond-lengths. Changing the parameter $c$, however also changes
the V1-V4 distance which effects the bare or direct $a_{1g}-a_{1g}$
hopping. We have carried out calculations for $c$=0, 1/3, 2/3 an 1.
The calculations for the hopping matrix elements are carried out
for the truly minimal set of $a_{1g}$ and $e_{g}^{\pi}$ {\it N}MTOs as
well as for the set where the O-$p$'s are kept active in addition
to V-$d$'s. This has been done to bring out the renormalization
effect coming from oxygen degrees of freedom. The results are shown
in FIG.\ \ref{distortion}. We see that the $pd$ contribution given by the difference
of direct $a_{1g}-a_{1g}$ and the renormalized $a_{1g}-a_{1g}$
hopping, almost vanishes for the ideal structure and increases
monotonically as the trigonal distortion increases 
towards to the value obtained in the real structure. The $pd$
contribution, apart from the case of ideal structure, is anti-bonding
whereas the direct (bare) $dd$ contribution is bonding. On top of
this, comes the even stronger trend that the bonding $dd$
interaction decreases with the V1-V4 bond distance, d, 
as $\approx$ d$^{-3.3}$. As a result, $a_{1g}-a_{1g}$ hopping
integral depends strongly on the distortion. As already mentioned,
in CNR's calculation, though the bare $dd$ hopping was obtained
following LCAO type of approach\cite{ashkenazi} on a correct V-V geometry
with correct V1-V4 bond distance, the trigonal distortion of the
VO$_{6}$ octahedra was assumed to be negligible with all V-O bond
lengths to be equal. This resulted into a direct $dd$ hopping of
-0.66 eV in good agreement with our corresponding estimate of -0.72
eV, while the $pd$ contribution gave rise to a small
renormalization of -0.06 eV. We note that the $pd$ contribution is
small and of bonding ({\it negative}) type as we obtained in our
calculation with ideal structure. This in turn proves the extreme
sensitivity of the hopping parameters on the correct geometry of
the system.
In FIG.\ \ref{ovl3} we show the comparison between
overlap of V1-$a_{1g}$ and V4-$a_{1g}$ {\it N}MTOs for the real and
ideal structures. Studying the figure, we notice that due to the
about 14 $\%$ reduction of V1-V4 distance in case of ideal
structure, the direct, bonding type $a_{1g}-a_{1g}$ overlap is much stronger
than compared to that in real structure which gives rise to the
strong slope observed in FIG.\ \ref{distortion}. Focusing on the
{\it pd} contribution, for the ideal structure, the $p-$ type tail
from V4 {\it N}MTO at O5 site (referred as $p_4$ in Eqn.2) 
passes almost through the node of the $a_{1g}$ orbital at V1 site, 
as shown with solid line, and gives rise to a negligible overlap
between $d_1$ and $p_4$ ({\it c.f.} Eqn.\ \ref{covalen}). For real
structure, on the other hand, the change of V1-O5-V4 angle causes
the $p_4$ tail having finite overlap with $a_{1g}$ orbital at V1
site which turn out to be positive (antibonding) in sign.

\subsection{Vertical pair model and validity of molecular orbital
  states}

In this sub-section, we discuss the consequences of the new set of
{\it N}MTO derived parameter values in the context of validity of     
vertical pair model and the molecular orbital formation.
   
Di Matteo {\it et. al.}\cite{matteo} analyzed in detail the parameter space of
the V$_{2}$O$_{3}$ problem in context of all possible orbital and
magnetic ground states configuration of the effective many-body
Hamiltonian by using variational procedure.  The orbital wave-functions
of the ferromagnetic state of the vertical
pair of V atoms at sites $a$ and $b$ were postulated to be given by:

\begin{equation}
\vert \psi_{\pm} \rangle_{ab} = 1/\sqrt{2} ( \vert \pm 1 \rangle_{a}
\vert 0 \rangle_{b} + \vert \pm 1 \rangle_{b} \vert 0 \rangle_{a})
\end{equation}

where $\vert  0 \rangle$ = $\vert e_{g}^{\pi},1 e_{g}^{\pi},2 \rangle$,
$\vert - 1 \rangle$ = $\vert a_{1g} e_{g}^{\pi},1 \rangle$ and
  $\vert 1 \rangle$ = 
$\vert a_{1g} e_{g}^{\pi},2 \rangle$
are three two-electron states constructed out of three
one-electron states, $e_{g}^{\pi},1$, $e_{g}^{\pi},2$ and $a_{1g}$.

The correlation energy of such a state was defined as the
difference between the ground state energy and its Hartree-Fock
approximation. 
Taking into account the hopping integrals in terms of reduced
 parameters, the ground state energy of such a pair is given by 
$ - \frac{(\mu - \rho)^{2}}{U-J}$, which involves back and forth
virtual hoppings out of $e_{g}^{\pi}$ state [$ -\frac{\mu^{2}}{U-J}$], that out 
of $a_{1g}$ state [$-\frac{\rho^{2}}{U-J}$] and the correlated hopping between
V atoms at $a$ and $b$ sites, where they exchange electrons in $a_{1g}$
and $e_{g}^{\pi}$ states simultaneously. U and J are Coulomb
and exchange integrals respectively. The latter mechanism
which arises due to {\it entangled} nature of the form of the
wave-function in (3) is absent in its Hartree-Fock approximation
and gives what Di Matteo {\it et. al.} termed as molecular correlation
energy. Two different regimes of solutions were defined depending
on the relative magnitudes of the correlation energy of the
ferromagnetic state of the vertical pair, {\it namely}, the molecular correlation energy and the
in-plane exchange energy (governed by the hopping
 processes in the basal plane):

(a) If the molecular correlation energy is larger than the in-plane
exchange energy then {\it the whole crystal consists of some
molecular units} and the variational wave-function should be
constructed in terms of molecular states given by (3).

(b) If the in-plane exchange energy is larger than the molecular
correlation energy, which tend to break the stability of the
correlated molecular states, one needs to construct the variational
wave-function in terms of single site atomic states.

Di Matteo {\it et. al.}\cite{matteo} approximated the in-plane exchange energy by
$\approx$ $\frac{(\alpha^{2} + \tau^{2})}{U-J}$ [considering the
hoppings along three 2NN bonds in the direction of 2, 3 and 3$^{'}$
and neglecting the hoppings involving reduced parameters $\beta$
and $\sigma$]

Considering the numerical values of $\mu, \rho, \alpha$ and $\tau$,
 as given by CNR and Di Matteo {\it et. al.} one gets:

\[
2 \mu \rho / (\alpha^{2} + \tau^{2}) \approx 4 - 3.5
\]

while using the parameters obtained by {\it N}MTO-downfolding one gets:

\[
2 \mu \rho / (\alpha^{2} + \tau^{2}) \approx .83
\]

[this ratio gets even smaller taking into account the hoppings
 related to $\beta$ parameter which is almost as large as $\tau$.]

Therefore, while the TB parameters given by CNR and Di Matteo {\it
et. al.} favor the formation of stable molecular orbital states,
the parameters provided by {\it N}MTO-downfolding procedure clearly do not
favor it. We therefore believe that it is very much needed
to repeat the calculations using the new set of hopping parameters
to shed light on the long standing puzzles in V$_{2}$O$_{3}$. This issue 
has been recently taken up in Ref.\cite{perkins}.

\subsection{Cr-doped V$_{2}$O$_{3}$ ((V$_{0.962}$Cr$_{0.038}$)$_{2}$O$_{3}$): 
the low-energy, tight-binding Hamiltonian}

The right, top panel of FIG. 9 shows the LDA band structure of 3.8$\%$ Cr doped
V$_{2}$O$_{3}$ in the paramagnetic insulating phase. Upon comparison with the
band structure of the undoped V$_{2}$O$_{3}$ in the paramagnetic metallic
phase, as presented in the left, top panel of FIG. 9, one finds that the
t$_{2g}$ bandwidth is reduced in the doped compound to about 2.25 eV from
about 2.5 eV in case of undoped compound. The right, bottom panel of FIG. 9
shows the a$_{1g}$ and e$_{g}^{\pi}$ bands switching off the a$_{1g}$-e$_{g}^{\pi}$
hybridization. The lattice expansion upon Cr doping, causes bottom of the
 a$_{1g}$ band to move up and the top of e$_{g}^{\pi}$ band to move down, giving
rise to max\{$\epsilon_{eg}$\} - min\{$\epsilon_{a1g}$\} $\approx$ 1.69 eV, in
comparison to max\{$\epsilon_{eg}$\} - min\{$\epsilon_{a1g}$\} $\approx$ 2.02 eV in
case of undoped compound.

The middle rows of Table I and II, lists the onsite energies and hopping matrix
elements corresponding to Cr doped V$_{2}$O$_{3}$. The crystal field splitting
is found to increase by 0.03 eV compared to undoped case. The magnitude of the 
dominant V1-V4 hopping is found to decrease from 0.51 eV in the undoped case to
0.43 eV in the doped case. These changes in the one-electron parameters were
found to be significant to drive the metal-insulator transition as explained
in Ref\cite{v2o3II}.

\subsection{Monoclinic V$_{2}$O$_{3}$: the low-energy, tight-binding
 Hamiltonian}

Finally, we thought it will be worthwhile to study the influence of the
monoclinic distortion in the low-temperature crystal structure on
the hopping integrals. It is of interest to know how much the
crystal structure change effects the hopping matrix elements. We
applied the same {\it N}MTO-downfolding machinery, described in
great detail in previous sections for V$_{2}$O$_{3}$ in corundum
structure. The LDA self-consistent potentials are generated by
TB-LMTO-ASA calculation with
potential sphere overlap less than 18 $\%$ and empty sphere overlap
less than 22 $\%$. The truly minimal $a_{1g}$ and $e_{g}^{\pi}$
basis sets are defined within the framework of the {\it
  N}MTO-downfolding technique for the monoclinic structure. For the
sake of comparison, we retained the corundum-symmetry-adopted
$a_{1g}$ and $e_{g}^{\pi}$ basis also in the monoclinic
structure. However, the further lowering of the symmetry in the
monoclinic phase introduces mixing between $a_{1g}$ and
$e_{g}^{\pi}$  orbitals at the same site, which is reflected as
crystal field terms in the on-site block of the real-space
Hamiltonian. The monoclinic distortion makes the various near
neighbor distances unequal\cite{struc_mono} compared to that in corundum phase.
While the vertical V1-V4 bond expands by 1.8 $\%$, the horizontal  
V1-V2 bond expands by about 4 $\%$ making the three, basal near
neighbor bond distances along 2, 3 and 3$^{'}$ unequal. It also makes the
bond distance along \underline{4} different from those along
\underline{8} and \underline{8'}, the bond distances along 5 and 
\underline{5} different from that of 6 and \underline{6} and, 6$^{'}$ and 
\underline{6$^{'}$}. The tight-binding hopping integrals and hopping elements computed as
elements of the orthonormalized $a_{1g}$ and $e_{g}^{\pi}$ {\it N}MTO
Hamiltonian are quoted in Table I and II. Focusing on the $a_{1g}-a_{1g}$
vertical pair hopping, we find that the value is further decreased
to -.44 eV compared to the value of -.51 eV in the corundum
structure. This reduction is primarily driven by the 1.8 $\%$
increase in the V1-V4 bond length and slight tilting of the V1-V4 
bond which decreases the magnitude of the bare $dd$ hopping
from -.72 eV in the corundum structure to -.64 eV in the monoclinic
structure. The $pd$ contribution due to integrated out oxygen tails
turned out to be +0.20 eV which can be compared with the value +.21
eV, that in corundum structure. The V1-O-V4 angle remains
essentially unaltered between the corundum and monoclinic
structure. 

Comparing the hoppings in other directions, the reduction is
maximum for the horizontal V1-V2 bond which expands by 4$\%$ over
the value in corundum structure. The other two near neighbor bonds
in the basal plane, 3 and 3$^{'}$ on the other hand contracts 
(the bond 3 by 0.7 $\%$ and 3$^{'}$ by .2 $\%$) which is reflected 
in the changes in hopping integrals. Similar distance dependent increase or
decrease can be observed for farther ranged hoppings.

In brief, though the low-temperature structural change induces
changes in the hopping parameters, these changes are not
drastic.  Till date, a direct, experimental evidence of orbital 
ordering is lacking and issue of orbital ordering, its existence
and type in V$_{2}$O$_{3}$ still remains highly controversial
\cite{oo}. Nevertheless, the role of orbital degrees of freedom
in stabilizing \cite{rice} the magnetic structure with broken 
trigonal symmetry of the corundum lattice remains to be plausible 
idea. In that case, the monoclinic distortion may possibly be the 
reflection of the peculiar spin and orbital ordering rather
than the cause for it.

\section{Summary and outlook}

To summarize, employing the {\it N}MTO-{\it downfolding} technique
and Wannier function representation of the Hamiltonian, we have
derived in a first-principles manner, the effective V-V hopping 
interactions corresponding to the low-energy, $t_{2g}$ bands of
V$_{2}$O$_{3}$. Our results show, contrary to popular believe, for
modeling of V$_{2}$O$_{3}$, inter-pair V-V hoppings are equally
important as V-V intra-pair hoppings. The significant changes in
hopping parameters compared to CNR parameters occur primarily due
to the neglect of trigonal distortion in the previous study and due
to the hopping processes via the integrated out e$^{\sigma}_{g}$
tails in addition to that via oxygen like tails, a fact not
considered before. This calls
the need for revisiting the many-body calculations, which start with
the assumption of vertical pairs as the building blocks.

The Wannier functions corresponding to the low-energy, t$_{2g}$
bands derived in this paper will serve as the basis to define
low-energy, multi-orbital Hubbard Hamiltonian for the LDA+DMFT
calculations for V$_{2}$O$_{3}$, which rely on the choice of
flexible, atom-centered, localized basis sets. Such calculations
have been already carried out. For details please see Ref\cite{v2o3II}. Considering the rather
delocalized nature of the real-space Hamiltonian of V$_{2}$O$_{3}$, 
it is quite natural to expect improvements on going beyond the single-site
approximation of DMFT and taking into 
account the cluster effect. For such study, it is crucial to decide 
on a minimal cluster which has the dominating effect and 
the vertical pair has been often discussed as a natural
choice. However, in view of {\it N}MTO-{\it downfolding} derived 
parameters and the breakdown of correlated molecular orbital like
states, the most tempting choice of the vertical pair as the cluster
seems to be hardly satisfactory. The cluster LDA+DMFT calculations with V1-V4 pair
show no qualitative difference with single site DMFT results
for corundum PI phase\cite{sasha}. The minimal cluster should include
in addition to vertical pairs V1 and V4, the near-neighbor V atoms
in the basal plane, V2, V3 and V3$^{'}$, which though is a computer
expensive DMFT job to carry out.

\newpage

\appendix
\section{{\it N}MTO method}

In the following, we describe the {\it N}MTO method which provides a tool
for direct generation of localized Wannier functions. The downfolding
procedure is also implemented in this framework to construct truly 
minimal basis sets which pick out selected bands.

In the {\it N}MTO method, a basis set of localized orbitals is constructed
from the exact scattering solutions for a superpositions, 
$\sum_{R}v_{R}(r_R)$, of short-ranged, spherically-symmetric potential
wells -- a so-called MT approximation to the potential. This is done
by first numerically solving the radial Schr\"{o}dinger's equations,
to find  $\varphi_{Rl}\left(\epsilon _{n},r_{R}\right)Y_{lm}\left({\bf \hat{r}}_{R}\right)$, 
the partial waves, for all angular momenta, $l,$ with
non-vanishing phase-shifts, for all potential wells, $R,$  and for a chosen set 
of energies spanning the region of interest, 
$\epsilon _{n}=\epsilon _{0},....,\epsilon _{N}$:
\[
-\left[ r\varphi _{Rl}\left( \varepsilon ,r\right) \right] ^{\prime \prime }=%
\left[ \varepsilon -v_{R}\left( r\right) -l\left( l+1\right) /r^{2}\right]
r\varphi _{Rl}\left( \varepsilon ,r\right)
\]

The partial-wave channels, $Rlm$ are partitioned into {\it active} and 
{\it passive} channels. The active channels are those for which one chooses
to have orbitals in the basis set, {\it i.e.} they are the chosen one-electron
degrees of freedom. The passive channels are said to be {\it downfolded}.

For each active channel, $\bar{R}\bar{l}\bar{m},$ a so-called 
{\em kinked partial wave} (KPW), 
$\phi _{\bar{R}\bar{l}\bar{m}}\left( \epsilon _{n},\mathbf{r}%
\right) ,$ is constructed. A kinked partial wave is basically a
partial wave with a tail joined continuously to it with a {\em kink} at a
central, so-called hard sphere of radius $a_{R}$. The tail of the kinked
partial wave is a {\em screened spherical wave}, 
$\psi _{\bar{R}\bar{l}\bar{m}}\left(
\varepsilon ,{\bf r}\right) ,$ which is essentially the solution with energy 
$\varepsilon $ of the wave equation in the interstitial between the hard
spheres, 
\[
-\Delta \psi \left( \varepsilon ,{\bf r}\right) =\varepsilon \psi
\left( \varepsilon ,{\bf r}\right) 
\]
with the boundary condition that,
independent of the energy, $\psi _{\bar{R}\bar{l}\bar{m}}\left( \varepsilon ,{\bf r}\right) $
go to $Y_{\bar{l}\bar{m}}\left( {\bf \hat{r}}_{R}\right) $ at the central hard sphere,
and to {\em zero} (with a kink) at all other hard spheres at the
neighboring sites. It is this latter 
{\em confinement,} which makes the screened spherical waves and
the KPWs localized when the energy is not too high. The default value of 
the hard-sphere radii, $a_{R},$ is 90 $\%$ of the appropriate covalent, 
atomic, or ionic radius. The above-mentioned boundary condition only applies to
the active components of the spherical-harmonics expansions of the screened
spherical wave on the hard spheres.  For the remaining {\it downfolded} or
{\it passive} components the screened spherical wave equals the corresponding
partial-wave solution of Schr\"{o}dinger's equation throughout the
MT-sphere, {\it i.e.} it has the proper phase shift. 

If one can now form a linear combination
of such kinked partial waves with the property that all kinks cancel, one
finds a solution of Schr\"{o}dingers equation with energy $\epsilon
_{n}.$ In fact, this kink-cancellation condition leads to the classical
method of Korringa, Kohn and Rostoker \cite{KKR} (KKR), but in a general
--so-called screened-- representation and valid for overlapping MT
potentials to leading order in the potential overlap. The screened KKR
equations are a set of energy-dependent, homogeneous linear equations, with
a matrix, $K_{\vec{R}\vec{l}\vec{m},\bar{R}\bar{l}\bar{m}}\left( \varepsilon
\right) ,$ whose rows and columns are labeled by the active channels. In the
{\it N}MTO method, we don't solve this set of secular equations, but proceed via
construction of energy-{\em in}dependent MTO basis sets which span the
solutions $\Psi _{i}\left( {\bf r}\right) $ with energies $\varepsilon _{i}$
of Schr\"{o}dinger's equation to within errors proportional to $\left(
\varepsilon _{i}-\epsilon _{0}\right) \left( \varepsilon _{i}-\epsilon
_{1}\right) ..\left( \varepsilon _{i}-\epsilon _{N}\right) ,$ where $%
\epsilon _{0},\epsilon _{1},...,\epsilon _{N}$ is the chosen 
{\em energy mesh } with N+1 points defined already. Such an 
energy-independent set of Nth-order MTOs is called an {\it N}MTO set.

The members of the {\it N}MTO basis set for the energy mesh 
$\epsilon _{0},...,\epsilon _{N}$ are superpositions, 
\begin{equation}
\chi_{\bar{R}\bar{l}\bar{m}}^{\left( N\right) }\left( {\bf r}\right)
=\sum_{n=0}^{N}\sum_{\vec{R}\vec{l}\vec{m}}\phi_{\vec{R}\vec{l}\vec{m}}
\left(\epsilon _{n},{\bf r} \right) L_{n \vec{R}\vec{l}\vec{m},
\bar{R}\bar{l}\bar{m}}^{\left( N \right)}
\end{equation}
of the kinked partial waves, $\phi _{\bar{R}\bar{l}\bar{m}}\left( \varepsilon ,{\bf r}\right) ,$
at the $N+1$ points (labeled by $n)$ of the energy mesh. Expression (1) is the
energy-quantized form of Lagrange interpolation,%
\[
\chi ^{\left( N\right) }\left( \varepsilon \right) \approx
\sum_{n=0}^{N}\phi \left( \epsilon _{n}\right) l_{n}^{\left( N\right)
}\left( \varepsilon \right) ,\quad l_{n}^{\left( N\right) }\left(
\varepsilon \right) \equiv \prod_{m=0,\neq n}^{N}\frac{\varepsilon -\epsilon
_{m}}{\epsilon _{n}-\epsilon _{m}},
\]
of a function of energy, $\phi \left( \varepsilon \right) ,$ by an $N$
th-degree polynomial, $\chi ^{\left( N\right) }\left( \varepsilon \right):$
The $N$th-degree polynomial, $l_{n}^{\left( N\right) }\left( \varepsilon
\right) ,$ is substituted by a matrix with elements, $L_{n\vec{R}\vec{l}\vec{m},
\bar{R}\bar{l}\bar{m}}^{\left( N\right) }\,,$ the function of energy, $\phi \left(
\varepsilon \right) ,$ by a Hilbert space with axes, $\phi _{\bar{R}\bar{l}\bar{m}}\left(
\varepsilon ,{\bf r}\right) ,$ and the interpolating polynomial, $\chi
^{\left( N\right) }\left( \varepsilon \right) ,$ by a Hilbert space with
axes, $\chi _{\bar{R}\bar{l}\bar{m}}^{\left( N\right) }\left( {\bf r}%
\right) .$

Note that the size of the {\it N}MTO basis is given by the number of active
channels and is independent of the number, $N+1,$ of energy points.
The energy-selective and localized nature of {\it N}MTO basis makes the
{\it N}MTO set flexible and may be chosen as truly minimal, that is, to span 
selected bands with as many (few) basis functions as there are bands. 
If those bands are isolated, the {\it N}MTO set spans the Hilbert space of the 
Wannier functions and the orthonormalized {\it N}MTOs are the
Wannier functions. But even if the bands of interest overlap other bands, 
it may be possible to pick out those few bands and their corresponding 
Wannier-like functions with the {\it N}MTO method. The {\it N}MTO
method can thus be used for direct generation of Wannier or
Wannier-like functions.

The Lagrange coefficients, $L_{n}^{\left( N\right) },$ as well as the
Hamiltonian and overlap matrices in the {\it N}MTO basis are expressed solely in
terms of the KKR resolvent, $K\left( \varepsilon \right) ^{-1},$ and its
first energy derivative, $\dot{K}\left( \varepsilon \right) ^{-1},$
evaluated at the energy mesh, $\varepsilon =\epsilon _{0},...,\epsilon _{N}.$
Variational estimates of the one-electron energies, $\varepsilon _{i},$
may be
obtained from the generalized eigenvalue problem,%
\[
\left( \left\langle \chi ^{\left( N\right) }\left| \mathcal{H}\right| \chi
^{\left( N\right) }\right\rangle -\varepsilon _{i}\left\langle \chi ^{\left(
N\right) }\mid \chi ^{\left( N\right) }\right\rangle \right) \mathbf{v}_{i}=%
\mathbf{0,}  \label{eq3a}
\]
with%
\[
\mathcal{H}\equiv -\Delta +\sum_{R}v_{R}\left( \left| \mathbf{r-R}\right|
\right) ,  \label{eq3b}
\]
or as the eigenvalues of the one-electron Hamiltonian matrix,%
\[
H^{LDA}=\left\langle \chi ^{\left( N\right) \perp }\left| \mathcal{H}\right|
\chi ^{\left( N\right) \perp }\right\rangle  \label{eq3}
\]
in the basis of \emph{symmetrically orthonormalized} {\it N}MTOs:%
\[
\left| \chi ^{\left( N\right) \perp }\right\rangle ~\equiv ~\left| \chi
^{\left( N\right) }\right\rangle ~\left\langle \chi ^{\left( N\right) }\mid
\chi ^{\left( N\right) }\right\rangle ^{-\frac{1}{2}}.  \label{eq4}
\]

In the present paper, the orbitals shown are {\it N}MTOs \emph{before}
orthonormalization because they are (slightly) more localized
than the orthonormalized ones. The hopping integrals and on-site elements
given in the tables are of course matrix elements of the \emph{%
orthonormalized} Hamiltonian.

For crystals, all calculations except the generation of the screened
structure matrix are performed in the Bloch $\mathbf{k}$-representation%
\[
\chi _{\bar{R}\bar{l}\bar{m}}^{\left( N\right) }\left( \mathbf{k,r}\right)
= 1/\sqrt{L} \sum_{T}\chi _{\bar{R}\bar{l}\bar{m}}^{\left( N\right) }\left( \mathbf{r-T}%
\right) \exp \left\{ 2\pi i\mathbf{k\cdot }\left( \mathbf{\bar{R}+T}\right)
\right\}
\]
where $T$ labels the $L (\rightarrow \infty)$ lattice translations and $\bar{R}$ the active sites in
the primitive cell. In order to obtain the orbitals and the Hamiltonian in
configuration space, Fourier-transformation over the Brillouin zone is
performed.

It is worth-mentioning here that this construction of a minimal {\it N}MTO basis 
set is different from standard L%
\"{o}wdin downfolding. The latter partitions a \emph{given}, large (say
orthonormal) basis into active $\left( A\right) $ and passive $\left(
P\right) $ subsets, then finds the downfolded Hamiltonian matrix as:%
\begin{eqnarray*}
\left\langle A\left( \varepsilon \right) \left\vert \mathcal{H}\right\vert
A\left( \varepsilon \right) \right\rangle & = & \left\langle A\left\vert \mathcal{%
H}\right\vert A\right\rangle \\
& & -\left\langle A\left\vert \mathcal{H}%
\right\vert P\right\rangle \left\langle P\left\vert \mathcal{H}-\varepsilon
\right\vert P\right\rangle ^{-1}\left\langle P\left\vert \mathcal{H}%
\right\vert A\right\rangle 
\end{eqnarray*}%
and finally removes the $\varepsilon $-dependence of the downfolded basis by 
\emph{linearizing} $\left\langle P\left\vert \mathcal{H}-\varepsilon
\right\vert P\right\rangle ^{-1}$ and treating the term linear in $%
\varepsilon $ as an overlap matrix. Obviously, since the {\it N}MTO set is exact
at $N+1$ energy points, it is more accurate.

Our present {\it N}MTO code is however not yet self-consistent, so we used the 
current Stuttgart tight-binding version of the linear-muffin-tin-orbital 
(TB-LMTO)\cite{lmto} code within the atomic sphere approximation
(ASA) to generate the LDA potentials. Despite this shape approximation for
the potential, the {\it N}MTO bands used in
the present paper are more accurate than LMTO bands, first of all because
the {\it N}MTOs do not use the zero-energy approximation in the interstitial
region and, secondly, because we use $N>1.$

\section{Computational details}

As mentioned in Appendix A, our present {\it N}MTO code is not self-consistent, we therefore 
used the current Stuttgart TB-LMTO-ASA code to generate the LDA potentials.
Such a potential in the atomic-spheres approximation is an overlapping
MT-potential, like the one handled by the {\it N}MTO method, but with the relative
overlaps, 
\begin{equation}
\omega _{RR^{\prime }}\equiv \frac{s_{R}+s_{R^{\prime }}}{\left\vert \mathbf{%
R}-\mathbf{R}^{\prime }\right\vert }-1,  \label{ovl}
\end{equation}%
limited to about 20\%. This limitation comes from the LMTO-ASA+cc method,
which solves Schr\"{o}dinger's equation by treating the overlap as a
perturbation (the so called combined-correction term, cc) \emph{and} uses
screened spherical waves of \emph{zero} kinetic energy in the $s$%
-interstitial. Poisson's equation is solved for the output charge density,
spherically symmetrized inside the \emph{same} atomic $s$-spheres. 

We now specify our computational set-up. 
The radii of the potential spheres, $s_{R},$
were dictated by our use of the LMTO-ASA method to generate the LDA
potentials. In order to limit the overlaps defined by equation (\ref{ovl}),
interstital --or empty-- spheres (E) were inserted in the non-cubic
structures. Table \ref{tables1_aB} gives the radii of the
potential spheres. As a result, the overlap between atomic spheres was 
$<$16\%, between atomic and empty spheres $<$18\%, and
between empty spheres $<$20\%.
We used the guidance given by the current version of the code in choosing 
the potential spheres appropriately.  

\begin{table}[h]
\caption{Radii $s_R$ of potential spheres in Bohr atomic units.}
\label{tables1_aB}
\begin{center}
{\setlength{\tabcolsep}{4pt} 
\begin{tabular}{cccccccccc} \hline
    & V    & O1   & O2  & E   & E1  & E2  & E3   & E4  & E5\\ 
    &      &      &     &     &     &     &      &     &\\ 
 PM & 2.46 & 1.88 &     & 2.36& 1.80&     &      &     & \\ 
 PI & 2.46 & 1.88 &     & 2.36& 1.80&     &      &     & \\ 
 AFI& 2.46 & 1.88 & 1.88& 2.36& 2.36& 1.80& 1.80 & 1.80&1.80  \\ \hline
\end{tabular}
}
\end{center}
\end{table}

\begin{table}[h]
\caption{LMTO basis sets
used in the self-consistent calculation of LDA potential.
$(l)$ means that the $l$-partial waves were downfolded within
in the LMTO-ASA+cc.}
\label{tables2_aB}
\begin{center}
\begin{textit}
{\setlength{\tabcolsep}{4pt} 
\begin{tabular}{ccllllllll}\hline
  & {\rm V} & {\rm O1} & {\rm O2} & 
     {\rm E} & {\rm E1} & {\rm E2} & {\rm E3} & {\rm E4} & {\rm E5}\\ 
  &  &  &  &  &  &  &  &  & \\ 
 {\rm PM} & spd & sp{\rm (}d{\rm )} &   & s{\rm (}pd{\rm )} & s{\rm (}p{\rm )}  &   &   &   &  \\ 
 {\rm PI} & spd & sp{\rm (}d{\rm )} &   & s{\rm (}pd{\rm )} & s{\rm (}p{\rm )}  &   &   &   &  \\  
 {\rm AFI}& spd & sp{\rm (}d{\rm )} & sp{\rm (}d{\rm )} & s{\rm (}pd{\rm )} & s{\rm (}pd{\rm )} &
 s{\rm (}p{\rm )}  & s{\rm (}p{\rm )}  & s{\rm (}p{\rm )}  & s{\rm (}p{\rm )}  \\ \hline
\end{tabular}
}
\end{textit}
\end{center}
\end{table}

The self-consistent valence-electron densities were
calculated with the LMTO bases listed in Table~\ref{tables2_aB}.
We found it is important to downfold the oxygen $d$
partial waves, rather than to neglect them (\textit{i.e.} to approximate
them by spherical Bessel functions when solving Schr\"{o}dingers equation,
and to neglect them in the charge density). 
Since the LMTO calculations were used
to produced the self-consistent charge densities, the energies, $\epsilon
_{Rl},$ for the linear $\phi _{Rl},\dot{\phi}_{Rl}$ expansions were chosen
at the centers of gravity of the \emph{occupied} parts of the respective DOS 
$Rl$-projections.

Finally, in the {\it N}MTO calculations, the hard-sphere radii, $a_{R},$ for the
active channels were chosen as 0.7$s_{R}.$ 

\begin{acknowledgments}
TSD gratefully acknowledges support from the MPG through the MPG-India partnergroup 
program. AIP thanks the Marie Curie grant MIF1-CT-2006-021820. 
The authors would like to acknowledge the hospitality of KITP, Santa Barbara
where the project was initiated.
\end{acknowledgments}


\begin{thebibliography}{99}
\bibitem{note} These are $e_{g}^{\sigma}$-s, in contrast to $e_{g}^{\pi}$-s
to be introduced later, but for simplicity we call them as  $e_{g}$. From
now on refering to $e_{g}$ will mean $e_{g}^{\sigma}$.
\bibitem{note1} In Matteheiss's paper the bands were plotted along
               the symmetry directions of the hexagonal BZ, where in the present
               paper they have been plotted along the symmetry
               directions of the rhombohedral BZ. We have also plotted
               the LMTO bands in the symmetry directions of the
               hexagonal BZ and they have been found to be in good
               agreement with that obtained by LAPW calculation of Matteheiss.
\bibitem{note_orb} The rather high contour chosen in the figure, tend to
  mask the antibonding $\langle  p_{1}, d_{4}
\rangle$ and $\langle  d_{1}, p_{4} \rangle$ hopping paths
and one erroneously focuses on the bonding $\langle  d_{1}, d_{4}
\rangle$ and $\langle  p_{1}, p_{4}
\rangle$ hopping paths. However, the $\langle  p_{1}, d_{4}
\rangle$ and $\langle  d_{1}, p_{4} \rangle$ gives larger
contribution than $\langle  p_{1}, p_{4}
\rangle$, and $\langle  p_{1}, d_{4}
\rangle$ + $\langle  d_{1}, p_{4} \rangle$ is antibonding (blue
  $p_1$ lob overlap with red $d_4$ lob and blue
  $p_4$ lob overlap with red $d_1$ lob).
\end{thebibliography}
\end{document}